%% file: J_Psi_DPPF_Arxiv.tex
\documentclass[preprint,aps,amsmath,superscriptaddress,nofootinbib]{revtex4}
\usepackage{color}
\input{rgb}
\usepackage{bm}
\usepackage{epsfig}
\usepackage{amsfonts}
\usepackage{bbm}

\def\nslash{n\!\!\!\slash}
\def\bnslash{\bar n\!\!\!\slash}

\def\dslash{\partial\!\!\!\slash}

\def\OMIT#1{}

\newcommand{\CH}[2]{\chi_{#1,#2}}

\newcommand{\bCH}[2]{\overline\chi_{#1,#2}}

\newcommand{\nn}{\nonumber} 

\newcommand{\bn}{{\bar n}}
\newcommand{\bea}{\begin{eqnarray}}
\newcommand{\eea}{\end{eqnarray}}

\newcommand{\bnP}{\bar {\cal P}}

\newcommand{\cP}{{\cal P}}

\newcommand{\beq}{\begin{equation}}
\newcommand{\eeq}{\end{equation}}
\newcommand{\SCETa}{\mbox{${\rm SCET}_{\rm I}$ }}
\newcommand{\SCETb}{\mbox{${\rm SCET}_{\rm II}$ }}
\newcommand{\SCETm}{\mbox{${\rm SCET}_{m}$ }}

\begin{document}


\preprint{\vbox{ \hbox{TUM-EFT 33/12} }}

\title{The Systematics of Quarkonium Production at the LHC and Double Parton Fragmentation  } 

\author{Sean Fleming\footnote{Electronic address: fleming@physics.arizona.edu}}
\affiliation{Department of Physics, 
         University of Arizona,
	Tucson, AZ 85721
	\vspace{0.2cm}}

\author{Adam K. Leibovich\footnote{Electronic address: akl2@pitt.edu}}
\affiliation{Pittsburgh Particle physics Astrophysics and Cosmology Center (PITT PACC)\\ Department of Physics and Astronomy, 
         University of Pittsburgh,
	Pittsburgh, PA 15260
	\vspace{0.2cm}}
	
\author{Thomas Mehen\footnote{Electronic address: mehen@phy.duke.edu}}
\affiliation{Department of Physics, 
	Duke University, 
	Durham,  NC 27708
	\vspace{0.2cm}}

\author{Ira Z. Rothstein\footnote{Electronic address: izr@andrew.cmu.edu}}
\affiliation{Department of Physics, 
        Carnegie Mellon University,
	Pittsburgh, PA 15213}
	
		\affiliation{California Institute of Technolgy, Pasadena, CA 91125}
	\vspace{0.2cm}\date{\today\\ \vspace{1cm} }


\begin{abstract}

In this paper we discuss the systematics of quarkonium production at the LHC.
In particular, we focus on the necessity to sum logs of the form $\log(Q/p_\perp)$ and $\log(p_\perp/m_Q)$.
We show that the former contributions are power suppressed, while the latter, whose contribution
in fragmentation is well known, also arise in  the  short distance (i.e., non-fragmentation) production mechanisms. 
Though these contributions are  suppressed by powers of $m_Q/p_\perp$,  they can be enhanced by inverse  powers of $v$, the relative velocity between heavy quarks in the quarkonium.
 In the limit $p_\perp \gg m_Q$ short distance production can be thought of as the fragmentation of a pair of partons (i.e., the heavy quark and anti-quark) into the final state quarkonium. We derive an all order factorization theorem for this process in terms of  double parton fragmentation functions (DPFF) and calculate the one-loop anomalous dimension matrix for the DPFF.

\end{abstract}

\maketitle

\newpage


Quarkonium production is a semi-inclusive hadronic observable that requires minimal non-perturbative input. Predictions for this observable depend only on  the usual parton distributions and a set of local non-perturbative quarkonium production matrix elements
that  can  be extracted from the data.  These predictions are based on the Non-Relativistic QCD (NRQCD) \cite{Bodwin:1994jh}
factorization theorems and are formulated as a double
expansion in $\alpha_s$ and $v$, where $v$ is the typical relative velocity of the heavy quarks in the bound state.

In this paper we will concentrate on the vector states $J/\psi$ and $\Upsilon$,
which have the quantum numbers $^{2S+1}L_J =  {}^3S_1$.  
At leading order in the $v$ expansion there is only one relevant NRQCD matrix element, which represents
the probability of a heavy quark-antiquark pair in a color- singlet ${}^3S_1$ state to form a quarkonium bound state. However,
 sub-leading contributions in the velocity expansion can receive kinematical enhancements \cite{frag} that scale
 as  powers  of $p_\perp/m_Q$.  Thus calculation of high $p_\perp$ quarkonium production should be formulated 
as  a systematic expansion in
 three parameters: $\alpha_s,v$, and $m_Q/p_\perp$. The relative importance of the various mechanisms depends on all three parameters. The parameters $\alpha_s(2 m_Q)$ and $v$ are fixed for a particular quarkonium state, but $p_\perp/m_Q$ clearly varies within the experiment depending on what $p_\perp$ is measured.  Currently, the LHC experiments CMS
 and ATLAS  have measured $J/\psi$ production in $pp$ collisions at $\sqrt{s} = 7$ TeV with $p_\perp$ as high as
 70 GeV~\cite{Chatrchyan:2011kc,Aad:2011sp}  and $\Upsilon$ production with  $p_\perp$ up to 24 GeV~\cite{Khachatryan:2010zg,Aad:2011xv} .
 
\begin{figure}
\begin{center}
\includegraphics[width=3.5in]{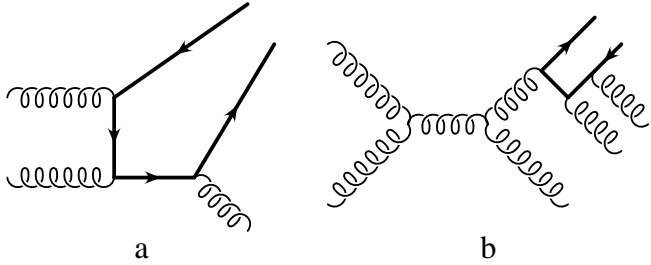}
\caption{a) The short distance production mechanism with the quark pair produced in a color-singlet state. b)  Gluon fragmentation production contribution to hadro-production. The two soft gluons emitted 
via E1 transitions are suppressed by $v^4$.}
\label{directandfrag}
\end{center}
\end{figure}

  The dominant contribution at asymptotically large  $p_\perp$
 will come from gluon fragmentation,  shown in Fig.~\ref{directandfrag}b.  In this paper we will follow the terminology of Ref.~\cite{Braaten:1993rw}
 and refer to direct production via non-fragmentation processes as short distance (SD) production.
 As will be explained below the fragmentation contribution naively scales as
 $(p_\perp^2/m_Q^2) v^4$ relative to SD production contributions shown in Fig.~\ref{directandfrag}a.
  The $v^4$ suppression  is due to the fact that in gluon fragmentation  the quark pair is produced in a color-octet state
 and color quenching requires  subsequent emission of soft  gluons which  vanishes in the static limit. 
 But, as will be shown below, this SD contribution is actually further suppressed by a factor of $(m^2_Q/p_\perp^2)$ at
 leading order in $\alpha_s(p_\perp)$. Thus the fragmentation contribution scales as $ (p_\perp^4/m_Q^4)v^4$ relative to LO SD production.
However,
 the NLO color-singlet SD contribution scales as $1/p_\perp^6$ and so fragmentation scales as  $(p_\perp^2/m_Q^2)(v^4/\alpha_s )$ relative to it. While this
 contribution is suppressed by an additional power of $\alpha_s(p_\perp)$ relative to the LO SD color-singlet piece, it does not fall off as steeply with $p_\perp$.
 Thus  at lower $p_\perp \sim 2 m_Q$ we expect the color-singlet SD  mechanism  to dominate, while at very high $p_\perp \gg 2 m_Q$ we expect fragmentation to dominate. The important question, however, is how each contribution affects the differential cross section in the intermediate $p_\perp$ region. 
 
 A  crucial distinction between high and low $p_\perp$ predictions is the size of the large logs that
 arise in perturbation theory. When the hierarchy
 \beq
 \label{hier}
 Q\gg p_\perp\gg m_Q
 \eeq
exists,  both $\log(Q/p_\perp)$ and $\log(p_\perp/m_Q)$ may appear in perturbation theory,
where $Q$ is the underlying hard scattering scale. The logs of $p_\perp/m_Q$ in single-parton
fragmentation can be resummed using  standard renormalization group (RG) techniques. This has already been accomplished
in the literature \cite{Braaten:1993mp,Braaten:1993rw,Braaten:1994xb}. However, the same type of logs arising in SD production have yet to be summed. 
These logs can be resummed by thinking of SD production as arising from double parton fragmentation (DPF) and then 
utilizing standard RG techniques~\cite{KQS}.  Of course, it is important to remember that the growth of
these logs is accompanied by a power suppression and thus we expect these logs to be
numerically important only in the intermediate, as opposed to asymptotic, regime.
Phenomenologically, most of data on $J/\psi$ and all data on $\Upsilon$ production at the Tevatron is at moderate $p_\perp$ where both single parton fragmentation and SD production mechanisms are important. This will continue to be true at the LHC, except for the very highest $p_\perp$ for $J/\psi$ production, where one is plausibly in the fragmentation regime.

 Let us determine the size of the logs in the intermediate regime where most of the available data is. Fragmentation
 and SD production are of the same order when\footnote{This is correct for comparing color-octet fragmentation to SD production. For ${}^3S_1$ color-singlet fragmentation, $v^4$ should be replaced with $\alpha_s^2$ but since numerically $v^2 \approx \alpha_s$ the estimate for $p_\perp/m_Q$ holds for this case as well.}
  \beq\label{ptest}
  \frac{p_\perp^4}{(2m_Q)^4} v^4 \sim 1 \qquad \textrm{(LO)} \qquad\qquad \frac{p_\perp^2}{(2m_Q)^2} \frac{v^4}{ \alpha_s} \sim 1 \qquad \textrm{(NLO)}.
  \eeq  
 For the $J/\psi \,(\Upsilon)$ system we will take $\alpha_s \sim v^2 \sim 0.3~(0.1)$.
Then both expressions in Eq.~(\ref{ptest}) yield  $p_{\perp}\sim 5\,(30)$~GeV. For these values of $p_\perp$, $\log(p_\perp/m_Q)$ is not a huge logarithm. However, for both $J/\psi$ and $\Upsilon$ the ratio $p_\perp/m_Q$ is comparable to the ratio $m_b/\Lambda_{\rm QCD}$. Resummation of $\log(m_b/\Lambda_{\rm QCD})$   is  required for accurate prediction in many processes involving heavy quarks.  It is likely such a resummation will be useful for quarkonium production as well. 
  
%
    
  An accurate prediction for this intermediate regime is important since the  norm of the production cross section is fixed by the NRQCD matrix elements, which  have to be extracted from the data. These matrix elements have been extracted from
production of $J/\psi$ at the $B$ factories~\cite{He:2009uf}, $e^+e^-$ annihilation at LEP \cite{Boyd:1998km}, photoproduction \cite{Amundson:1996ik},   $B$ decays \cite{Beneke:1998ks,Ma:2000bz}, as well as from fixed target \cite{Beneke:1996tk} and hadroproduction~\cite{CL1,CL2}.
   For a recent analysis that extracts  NRQCD matrix elements from a global fit to a wide range of experiments using NLO theoretical calculations, see Refs.~\cite{Butenschoen:2012qh,Butenschoen:2011yh}.   The extraction in Refs.~\cite{CL1,CL2} involves interpolating between (non-resummed) SD production and fragmentation and   it is unclear how to estimate the errors involved in this process given the merger regime is contaminated
  by the aforementioned large logs. Thus the accuracy of the extraction will be enhanced
  by the log resummation studied here. As such,  this paper can be thought of as a continuation of the study
  started in \cite{CL1,CL2} at leading logarithmic order.
   
 In the asymptotic regime where 
  \beq
  \frac{p_\perp^4}{(2m_Q)^4} v^4\gg 1,
 \eeq
  fragmentation dominates. A classic prediction of NRQCD in this regime is that quarkonium production is dominated by color-octet fragmentation and the quarkonium is produced with purely transverse polarization \cite{CW,BR, Leibovich:1996pa}.  Presently, there is no indication of this  trend in the data \cite{data}.   Currently available data does not probe the asymptotic regime for the $\Upsilon$ system, but the failure of the prediction for $J/\psi$ calls into question the validity of the NRQCD power counting in charmonium.    In the $J/\psi$ one must recall that treating this system as  Coulombic is questionable and might  require a power counting distinct from NRQCD \cite{B,FLR}.  Since much of the data falls in the intermediate regime we expect the resummation of $\log(p_\perp/m_Q)$ to shed some light on the polarization puzzle.

  The purpose of this paper is to derive factorization theorems and evolution equations
that will make it possible to resum $\log(p_\perp/m_Q)$. We will also show that contributions with $\log(Q/p_\perp$), where $Q\gg p_\perp$, are power suppressed.  Resummation of $\log(p_\perp/m_Q)$ requires the introduction of the power suppressed
  double parton fragmentation function (DPFF), which was first introduced in Ref.~\cite{KQS}. We
  will show that the former types of logs are suppressed by powers of $p_\perp/Q$.
  We will use soft-collinear effective theory (SCET) \cite{Bauer:2000ew,Bauer:2000yr}
  for our derivation of factorization.\footnote{Like all SCET based factorization, our result relies upon
 standard methods of proof \cite{CSS} when it comes to the cancellation of the so-called Glauber
 contributions.} In Appendix A we review the minimal SCET
  formalism and notation used in this paper. 
 We do not explicitly perform the resummations in this paper, leaving that for
 a subsequent publication. The new results in this paper include the necessary
 factorization theorems for both types of resummations as well  the one-loop
 anomalous dimension matrix for the DPFF.

 \section{Factorization}
 The inclusive differential cross section for the production of a quarkonium state $H$ with mass $M_H$, four-momentum $p$ and transverse momentum $p_\perp$ via the collision of two incoming hadrons, $h_1$ and $h_2$, with momentum $p_1$ and $p_2$, respectively,  is:
\bea
\label{qcdcs}
\frac{d \sigma}{dp^2_\perp}(h_1 + h_2 \to H +X)&=&\frac{1}{2 (p_1+p_2)^2} \,
 \frac{1}{4} \sum_{spins} \sum_X \int \frac{d y}{(4 \pi)^2} (2\pi)^4 \delta^4(p_1+p_2 -p-p_X) \nn \\
 & & \times  |{\cal M}( h_1 h_2 \to H(y,p_\perp) + X)|^2 \, .
\eea
Here the azimuthal angle of $H$ is integrated over, and $y$ is the rapidity of $H$, which is restricted to be in the range $-2.4 \lesssim y \lesssim 2.4$ in the LHC experiments~\cite{Chatrchyan:2011kc, Aad:2011sp}.  
The hadronic matrix element above receives contributions from all scales between  $Q$, the invariant mass of
the partonic collision,  and the hadronic scale $\Lambda\sim 1\, \textrm{GeV}$.  In addition to these scales
there are two other relevant scales that may be hierarchically separated from $Q$ and $\Lambda$, namely
$p_\perp$ and the heavy quark mass $m_Q$. In principle, perturbation theory could be
plagued by large logs of the ratios of these scales.  Thus to be able to calculate within a well-defined
approximation scheme, we need specify which regime we are in. We need to consider the two possible hierarchies
\beq
({\rm I})~~Q\gg p_\perp \gg m_Q,
\eeq
and 
\beq
({\rm II})~~Q\sim p_\perp \gg m_Q.
\eeq

We will now show that the regime (I) is power suppressed. In this region, the physical picture is that 
the quarkonium must be accompanied by at least two nearly back-to-back jets  whose {\it net} $p_\perp\ll Q$, 
and whose {\it  total} invariant mass is $\sim Q$. 
Let us consider  the appropriate theory  below the scale $Q$. 
In the lab frame the incoming hadrons move along the $z$ axis and their light-cone momenta $(k^+,k^-,k_\perp)$ scale as
$(\sqrt{s}, 0, 0)$ and $(0,\sqrt{s}, 0)$. Since the quarkonium's $p_\perp$ is the infrared scale, the initial-state radiation collinear to the incoming beams has four-momenta scaling as
\begin{equation}
p_n \sim (Q , \frac{p_\perp^2}{Q }, p_\perp) \qquad \bar p_{\bn} \sim (\frac{p^2_\perp}{Q}, Q , p_\perp) \,.
\end{equation}
In addition there exists soft radiation whose momentum scales as
\beq
p_s \sim (p_\perp,p_\perp,p_\perp).
\eeq
Taking as our expansion parameter $\lambda\equiv p_\perp/Q$ we see that while the collinear momenta scale
as $Q(1,\lambda^2,\lambda)$ the soft momenta scale as $Q(\lambda,\lambda,\lambda)$. This is the scaling associated
with  \SCETb\!\!, since  soft modes have large enough momenta to change the $p_\perp$ components of the collinear modes, 
as opposed to \SCETa\!\!, where the soft (in this context often called ultrasoft) scale as $(\lambda^2,\lambda^2,\lambda^2)$. 
Thus at the scale $Q$ the hard central jets are integrated out, and we match onto \SCETb where the infrared scale is $p_\perp$ and the
quark mass is irrelevant (unless we are interested in corrections of order $m_Q/Q$, which we will
ignore).

\begin{figure}
\begin{center}
\includegraphics[width=3.5in]{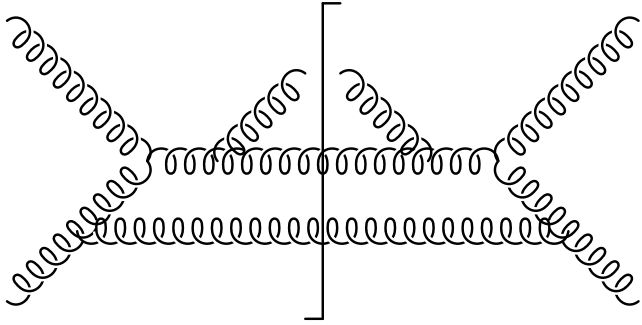}
\caption{A  typical contribution to gluon fragmentation with two hard central jets.
The lines going across the cut are integrated out, while the uncut lines represent the
fragmenting gluon.}
\label{central}
\end{center}
\end{figure}

Now that we have established the problem is posed in \SCETb\!\!,  the proof of power suppression is identical to
the proof given in Ref.~\cite{Chiu:2012ir} for the case of Higgs production at $p_\perp\ll m_{Higgs}$. 
The suppression of central jets arises for both single and double parton fragmentation, as the argument runs the same way
in both cases: the two (or more) hard partons corresponding to the central jets cross the cut
and are integrated out at the high scale as depicted in Fig.~\ref{central}. The power suppression
of such a contribution can be seen by noting that when there are no central jets the leading 
contribution scales as $1/p_\perp^2$. This scaling arises either due to a collinear emission or
virtual emission, which scales as $\delta^2(p_\perp)$.  When all of the momentum crossing
the cut is hard, such a scaling factor is necessarily absent and thus the resulting operator
which is generated is necessarily power suppressed. The situation with both central
jets and radiation down the beam pipe (though still satisfying $p_\perp\gg m_Q$) is reproduced
by the effective theory via the one loop matrix element of the power suppressed operator
generated by the central jets.

Let us now consider the regime (II).
When $p_\perp \sim Q$  the scale that controls the IR physics
is $m_Q$ and the scaling parameter is $\lambda \equiv m_Q/p_\perp$. Furthermore, we wish to keep the
full dependence on the quark mass so we match onto \SCETm~\cite{Rothstein:2003wh,Leibovich:2003jd} at the scale $Q$.
In this particular case, as we shall see, there is no contribution from soft radiation at all, so
the distinction between an \SCETa and \SCETb type theory becomes moot. 
 The light-cone components of the quarkonium momentum scale as
\begin{equation}
 p \sim p_\perp \bigg(\sqrt{1+\frac{m_H^2}{p_\perp^2}} e^y , \sqrt{1+\frac{m_H^2}{p_\perp^2}} e^{-y},\hat{n}_{\perp} \bigg) \sim 
 p_\perp \big(e^y, e^{-y},\hat{n}_{\perp}\big)\,,
\end{equation}
so 
\begin{equation}
p \approx \frac{\bn'\!\cdot\!p}{2}n'^{\mu}
 \qquad\textrm{with}\qquad n'^{\mu} = (1,\frac{\hat{n}_{\perp}}{\cosh y},\tanh y) \,,
\end{equation}
where $n'^{\mu}$ is given in standard four-component notation, $n'\cdot \bn'=2$, $\bn'\cdot p = 2 m_\perp \cosh y$, $m^2_\perp = p_\perp^2 + M^2_H$, and $\hat{n}_\perp$ is a unit vector in the $\perp$ direction.
We distinguish the direction perpendicular to the beam direction, denoted by $\perp$, from the direction perpendicular to the quarkonium direction, denoted by $\perp'$. Corrections to these leading terms are suppressed by powers of $m_Q/p_\perp$. 
The invariant mass of the final state remnants $X$ in the production process is $p^2_X = [(p_1+p_2)-p]^2 = 4E_\textrm{cm}(E_\textrm{cm}-m_\perp \cosh y)+ M_H^2 \approx 4E_\textrm{cm}(E_\textrm{cm}-p_\perp \cosh y) $. We restrict ourselves to the regime where $E_\textrm{cm} \gtrsim p_\perp \sim \sqrt{p^2_X} \gg M_H$, so that the final state remnant can be integrated out at the hard scattering scale $Q \sim p_\perp$.\footnote{
There is an additional scaling  that we could consider.   If $E_\textrm{had} \gg \hat{E}_\textrm{partonic}$ then we are in the regime where the parton light-cone momentum fraction $x$ goes to zero (a.k.a.\ small $x$), and the existence of terms scaling like $\ln x$ would be of concern. We will not consider such a scenario since it goes beyond the scope of this work.} 

Since the remnants of the production process have an invariant mass $p^2_X\sim Q^2$, they must be integrated out when matching onto \SCETm\!\!. This is done in the manner of an operator product expansion where the differential cross section in Eq.~(\ref{qcdcs}) is expanded in terms of all the SCET operators allowed by the symmetries of the theory. 
We must consider contributions from SCET operators with light quarks and gluons that can produce heavy quarks through  insertions of the \SCETm Lagrangian. SCET operators that involve only light quarks, gluons, or at most one heavy quark correspond to standard fragmentation, whereas operators with a heavy quark--anti-quark bilinear give rise to SD production. 
Since the fields in SCET have a power counting associated with them only a limited number of operators arise at each order in the power counting with corrections suppressed by $\lambda \sim m_Q/p_\perp$. The fragmentation and SD production contributions are of different order in the SCET power counting, so we can consider each in turn. 

The generic form of the SCET operators that are required  is
\begin{equation}
{\cal O}=O_{n}  O_{\bn}O_{n'}  {\cal P}^H_{n',Q} O_{n'}  \,, 
\end{equation}
where 
\begin{equation}
{\cal P}^H_{n', Q}=\sum_{X_n'} | H_{n',Q} + X_{n'}\rangle \langle H_{n',Q} + X_{n'}|\,.
\end{equation}
Here $H_{n',Q}$ is a quarkonium state that has a large light-cone momentum component $Q$ in the $n'$ direction, and $X_{n'}$ are states that are also collinear in the $n'$ direction. The operators $O_{n}$ and $O_{\bn}$  include fields in the $n$ and $\bar{n}$ directions respectively that are collinear to the initial state,  and $O_{n'} $ contains fields in the $n'$ direction that will eventually hadronize into a jet that includes the quarkonium state.

\begin{figure}
\begin{center}
\includegraphics[width=6in]{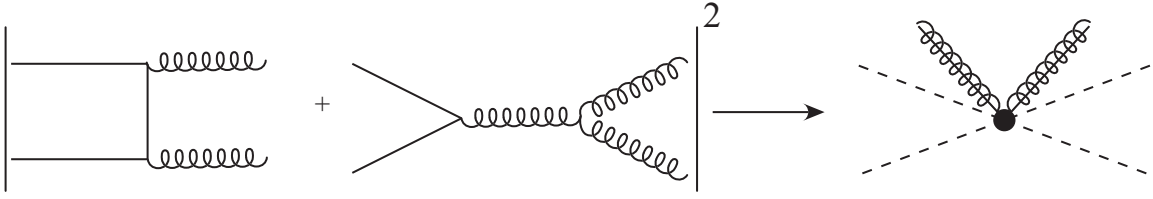}
\caption{Matching $O^{G}_{qq}$ at leading order. On the left is the square of the full theory amplitudes, and on the right is the matrix element of $O^{G}_{qq}$. At this order the matching coefficient is proportional to $\alpha_s^2(Q)$.}
\label{GqqMatching}
\end{center}
\end{figure}

Given the four possible initial parton combinations (schematically $qq, qG, Gq, GG$), there are twelve types of fragmentation operators (fragmenting of a light quark, heavy quark, or gluon) and four types of SD operators. Since the steps we take to arrive at the final factored form for the matrix element of each operator are the same, we will consider one operator of each type in detail.
The results are easily generalized to the other situations. The fragmentation operator for an incoming $q\bar q$ to produce an outgoing gluon in the $n'$ direction is
\bea
\label{fragcontGqqex}
{\cal O}^{G}_{qq}&=& \int d \omega_i  d \bar \omega_j  d \omega'_k  C(\omega_i,\bar \omega_j, \omega'_k)
(\bar \chi_{n, \omega_2}\frac{\bnslash}{2} \chi_{n,  \omega_1})
(B^{\nu A }_{n^\prime, \omega'_1}
 {\cal P}^H_{n',Q} 
B^{ \rho A }_{n^\prime, \omega'_2})
(\bar \chi_{\bn, \bar \omega_2} \frac{\nslash}{2}\chi_{\bn,  \bar \omega_1}) \,, 
\eea
where $i,j,$ and $k$ run from one to two.  This operator is arrived at after Fierzing the full theory diagram.
We  have kept only the contributions that lead to a non-vanishing matrix element at leading power.
This operator scales as $\lambda^6$ and can produce a $Q\bar Q$ pair through a time ordered product with an $O(\lambda^0)$ interaction term from the \SCETm Lagrangian.  The hard matching coefficient $C(\omega_i,\bar \omega_j, \omega'_k)$ is determined by perturbatively matching this operator onto the full theory and is therefore given by an expansion in  $\alpha_s(Q)$. For example, the matching of ${\cal O}^{G}_{qq}$ is depicted in Feynman diagrams in Fig.~\ref{GqqMatching}. The matching coefficient at tree level is proportional to $\alpha_s^2(Q)$.  However, since the production of a $Q\bar Q$ requires the insertion of an interaction term from the \SCETm Lagrangian, this operator will acquire an additional $\alpha_s(2m_Q)$. As a result the fragmentation contribution is proportional to $\alpha^2_s(Q) \alpha_s(2m_Q)$.

\begin{figure}
\begin{center}
\includegraphics[width=6in]{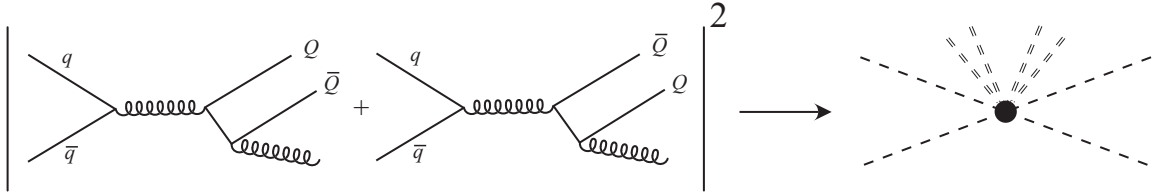}
\caption{Matching of $O^{QQ}_{qq}$ onto the full theory at leading order. On the left are two leading order Feynman diagrams that contribute to the production of a  $Q\bar{Q}$ pair from an incoming $q \bar q$ pair . On the right is the tree-level matrix element of  $O^{QQ}_{qq}$. The dashed lines are incoming and outgoing collinear light quark lines, and the dashed double lines are incoming and outgoing heavy quarks. At this order the matching coefficient is proportional to $\alpha^3_s(Q)$. }
\label{QQqqmatching}
\end{center}
\end{figure}

The SD production operator that describes the production of a $Q\bar Q$ pair from an initial light $q\bar q$ is
\bea
\label{fragcontQQqqex}
{\cal O}^{Q\bar{Q}}_{qq}&=&\sum_a \int d \omega_i  d \bar \omega_j  d \omega'_l  C^a(\omega_i,\bar \omega_j, \omega'_l) \\
&&\times
(\bar \chi_{n, \omega_2}\frac{\bnslash}{2} \chi_{n,  \omega_1})
(\bar \chi_{n^\prime, \omega'_2}\Gamma^{a(\nu)} \{1,T^A\}  \chi_{n^\prime,  \omega'_1} {\cal P}^H_{n',Q} 
\bar \chi_{n^\prime,\omega'_3}\Gamma^{a}_{(\nu)} \{1,T^A\} \chi_{n^\prime, \omega'_4})
(\bar \chi_{\bn, \bar \omega_2} \frac{\nslash}{2}\chi_{\bn,  \bar \omega_1}) \,, \nn 
\eea
where $l = 1, \dots, 4$, and $\Gamma^{a(\nu)} \in \frac{1}{2}\{\bnslash',\bnslash' \gamma_5, \bnslash' \gamma^\nu_{\perp'}\}$ with
$\gamma^\nu_{\perp'}= \gamma^\nu -n'^\nu \bnslash'/2 - \bn'^\nu \nslash'/2$.
Note the heavy quark--anti-quark pair can be in either a color-singlet or color-octet state.
This operator scales as $\lambda^8$ in the \SCETm power counting and is therefore $\lambda^{2} \sim m_Q^2/p_\perp^2$ suppressed relative to the fragmentation contribution. As for the fragmentation contribution, the hard matching coefficients $C^a$ are determined by perturbatively matching this operator onto the full theory.  For example, the lowest order matching, shown in  Fig.~\ref{QQqqmatching}, is proportional to $\alpha_s^3(Q)$.

We pause here to discuss a point made in the introduction; namely that at leading order in $\alpha_s$
the color-singlet ${}^3S_1$ SD  contribution scales as $m_Q^4/p_\perp^4$ relative to gluon fragmentation.
This is in apparent contradiction with our operator
analysis which leads us to conclude that the suppression of DPF is only of order $\lambda^2 \sim m_Q^2/p_\perp^2$. 
The resolution is that at leading order in $\alpha_s$
the color-singlet ${}^3S_1$ SD contribution does not have a leading in $\lambda$ DPF component, in other words the 
matching of the color-singlet ${}^3S_1$ SD contribution onto the leading DPFF vanishes at leading order in $\alpha_s$.
However, there are power corrections to the DPFF that scale as $\lambda^4$ (and higher) relative to the gluon fragmentation function.
Such sub-leading DPFFs could for example have the form of the leading DPFF with factors of the SCET covariant derivative inserted or 
have explicit factors of the quarkonium mass. It is these subleading DPFF contributions onto which the color-singlet ${}^3S_1$ SD contribution
matches at leading order in $\alpha_s$. However, at NLO in $\alpha_s$, the color-singlet ${}^3S_1$ SD contribution has a 
contribution whose scaling is $m_Q^2/p_\perp^2$ suppressed relative to fragmentation and which matches onto a leading DPFF.


Next we take matrix elements of the SCET operators between $ p$ states along the $n$ and $\bn$ directions.
The usoft fields decouple  from the collinear fields in the action by using the BPS field redefinition \cite{Bauer:2001yt}, which decouples the Hilbert spaces of the various modes and allows us to factorize the matrix element. After these steps the matrix element of the  fragmentation operator in Eq.~(\ref{fragcontGqqex}) is
\bea
\label{fragcontGqqfierzex}
\langle p_n p_\bn | {\cal O}^{G}_{qq} | p_n p_\bn \rangle &=&  \int d \omega_i  d \bar \omega_j  d \omega'_k  
C(\omega_i,\bar{\omega}_j, \omega'_k)\langle p_n |\bar \chi_{n, \omega_2}  \frac{\bnslash}{2} \chi_{n,\omega_1} | p_n \rangle
\langle  p_\bn | \bar \chi_{\bn, \bar{\omega}_2}  \frac{\nslash}{2}  \chi_{\bn, \bar{\omega}_1}|p_\bn \rangle \nn \\&\times&
\langle 0 |  \bigg[B^{ A \mu }_{n^\prime,\omega'_1}Y^{BA}
  {\cal P}^H_{n',Q}
Y^{AC}B^C_{n^\prime,\omega'_2\, \mu}\bigg]|0\rangle.
\eea
where $Y^{AB}$ are Wilson lines in adjoint representation  that run along the light-cone from infinity
to the point where the operator is situated, which we take to be the origin.
Since the projection operator is limited to the collinear sector and the soft Wilson
lines end at the same point as a consequence of the multipole expansion,
the $Y$'s cancel. Note that each state scales as $\lambda^{-1}$ and the projection operator ${\cal P}^H_{n',Q}$ scales as $\lambda^{-2}$ so the final matrix element of the fragmentation operator scales as $\lambda^0$.
As mentioned previously, it is easy to generalize this to the other initial and final states in SCET.

Next we consider the matrix element of the SD production operator in Eq.~(\ref{fragcontQQqqex}). After factoring usoft from collinear, this matrix element is
\bea
\label{fragcontQQqqfierzex}
\langle p_n p_\bn | {\cal O}^{Q\bar{Q}}_{qq} | p_n {p}_\bn \rangle &=& \sum_a  \int d \omega_i  d \bar \omega_j  d \omega'_l 
C^a(\omega_i,\bar{\omega}_j, \omega'_l)\langle p_n |\bar \chi_{n, \omega_2}  \frac{\bnslash}{2} \chi_{n,\omega_1} | p_n \rangle
\langle  {p}_\bn | \bar \chi_{\bn, \bar{\omega}_2}  \frac{\nslash}{2}  \chi_{\bn, \bar{\omega}_1}|{p}_\bn \rangle \nn \\&\times&
\langle 0 | \bar \chi_{n^\prime, \omega'_2}\Gamma^{a(\nu)} \{1,T^A\} \chi_{n^\prime,  \omega'_1} {\cal P}^H_{n',Q} 
\bar \chi_{n^\prime, \omega'_3}\Gamma^a_{(\nu)}\{1,T^A\}  \chi_{n^\prime, \omega'_4}|0\rangle,
\eea
where again the soft Wilson lines cancel.  The matrix element scales as $\lambda^2$ so is suppressed relative to the fragmentation matrix element.

The matrix elements involving the incoming states $|p_n\rangle$ and $|p_\bn \rangle$ in Eqs.~(\ref{fragcontGqqfierzex},\ref{fragcontQQqqfierzex}) are related to the parton distribution functions (PDFs)~\cite{Bauer:2002nz}:
\bea
\label{quarkpdf}
\frac{1}{2} \sum_\textrm{\small spin} \langle p_n(p))| \bar \chi_{n,\omega_1} \bnslash \chi_{n,\omega_2} |p_n(p)\rangle
& = & 4 \bn\cdot p \int^1_0 dz \, \delta(\omega_-) \delta(\omega_+ - 2 z \bn\cdot p) f_{i/p}(z)\\
&-& 4 \bn\cdot p \int^1_0 dz \, \delta(\omega_-) \delta(\omega_+ + 2 z \bn\cdot p)  f_{\bar i/p}(z),\nn \\
\frac{1}{2} \sum_\textrm{\small spin} \langle p_n(p)| \textrm{Tr}\big[ B^\mu_{n,\omega_1} B^{n,\omega_2}_\mu \big] |p_n(p)\rangle
& = & -\frac{ \omega_+ \bn\cdot p}{2} \int^1_0 dz \, \delta(\omega_-) \delta(\omega_+ - 2 z \bn\cdot p) f_{g/p}(z) \,, \nn
\eea
where $\omega_\pm = \omega_1\pm \omega_2$, $ f_{i/p}(z)$ is the quark PDF, $ f_{\bar i/p}(z)$ is the anti-quark PDF, and $f_{g/p}(z)$ is the gluon PDF. Futhermore, the vacuum matrix element of the fragmentation operator in Eq.~(\ref{fragcontGqqfierzex}) can be related to the standard fragmentation function that gives the probability of finding in the gluon a quarkonium state $H$ moving in the $n'$ direction with large light-cone momentum $\bn'\cdot p$:
\bea
\label{gluefrag}
& &
\frac{1}{N^2_c-1}\langle 0| 
 \textrm{Tr}\big[ B^{\mu }_{n^\prime,\omega'_1}
 {\cal P}^H_{n',\bn'\cdot p}
 B_{n^\prime,\omega'_2\, \mu}\big]
|0\rangle\\
& & \hspace{20 ex} = -\frac{4}{ \omega'_+}
\int_0^1\frac{dz}{z} \delta (\omega'_-) \, \delta\!\left(z  -\frac{ 2 \bn'\cdot p }{\omega'_+} \right)D_{H/g}(z)\nn \,.
\eea
For completeness we give the SCET definition of the light-quark to quarkonium fragmentation function,
\bea
\frac{1}{2N_c} \textrm{Tr} \langle 0| 
\bnslash' \chi_{n',\omega_1'} 
 {\cal P}^H_{n',\bn'\cdot p}
\bar\chi_{n',\omega_2'}
|0\rangle
& = &2
\int_0^1\frac{dz}{z} \delta (\omega'_-) \, \delta\!\left(z  - \frac{2 \bn'\cdot p }{\omega'_+} \right)D_{H/q}(z) \,. 
\eea
These definitions agree with those in Refs.~\cite{Collins:1981uw,Procura:2009vm,Jain:2011xz}.
Substituting Eqs.~(\ref{quarkpdf}, \ref{gluefrag}) into Eq.~(\ref{fragcontGqqfierzex}) we arrive at the familiar factored form for the fragmentation cross section in proton-proton collisions:
\begin{equation}
(4 \pi)^2\frac{d^2\sigma}{d p^2_\perp dy} = \int d x_1 d x_2 \frac{d z}{z} \,  \hat \sigma(x_1, x_2, z, p_\perp,y) f_{q/p}(x_1) f_{\bar{q}/p}(x_2) D_{H/g}(z)\,,
\end{equation}
where $\hat\sigma$ is the short-distance partonic differential cross section for producing a gluon from the collision of a quark and anti-quark.

Double parton fragmentation is the kinematic situation in which a collinear, highly energetic, nearly on-shell heavy quark--anti-quark pair hadronizes into an energetic  quarkonium.  In every quarkonium production process, the heavy quark and heavy anti-quark have small relative momenta, so  they are  collinear to each other. What distinguishes double parton fragmentation from, for example,  threshold production, is the large boost the heavy quark--anti-quark pair have relative to the lab frame. Because $p_\perp \gg m_Q$ we can think of the heavy quark and anti-quark as light-like collinear SCET modes.  For this to have an invariant meaning, the heavy quark--anti-quark pair must be recoiling against one or more energetic jets of partons, so there exists a  scale in the problem much larger than $m_Q$. The energetic partons (quarks) will form a jet via collinear radiation and generate a set of large logs that would not be present if the quark pair was produced
nearly at rest in the lab frame.

Therefore, the vacuum matrix elements in the SD production operators are also fragmentation functions, but of a new type. The DPFF is defined in terms of the matrix element in Eq.~(\ref{fragcontQQqqfierzex}) by
\bea
\label{QQfragfun}
& & \langle 0| 
\bar \chi_{n^\prime, \omega'_2}   \Gamma^{a (\nu) } \{ \mathbbm{1},T^A\} \chi_{n^\prime,  \omega'_1}
 {\cal P}^H_{n',\bn'\cdot p}
\bar \chi_{n^\prime,  \omega'_4}   \Gamma^a_{ (\nu) } \{ \mathbbm{1},T^A\}  \chi_{n^\prime,  \omega'_3} |0\rangle
\\
& & =  8\, \delta(\omega'_1-\omega'_2+\omega'_3-\omega'_4) \int \frac{dz}{z} \, du \, dv \,\delta(z-\frac{\bn'\!\cdot\! p}{\omega'_1-\omega'_2})
 \delta(v-1-z\frac{\omega'_2}{ \bn'\!\cdot\! p}) \delta(u-z\frac{\omega'_4}{ \bn'\!\cdot\! p})\nn\\
& &\hspace{30ex}\times z D_{a\{ 1,8\}}^{Q\bar Q}(u,v,z) \nn \,.
\eea
This distribution is a combination of fragmentation function and light-cone distribution amplitude. 
The light-cone momentum fraction variables are
\bea
\label{lcmfcv}
z &=& \frac{\bn'\!\cdot\! p}{\omega'_1-\omega'_2}=\frac{\bn'\!\cdot\! p}{\omega'_4-\omega'_3}\\
v &=& z\frac{\omega'_1}{ \bn'\!\cdot\! p}=1+z\frac{\omega'_2}{ \bn'\!\cdot\! p}\nn\\
u &=& z\frac{\omega'_4}{ \bn'\!\cdot\! p}=1+z\frac{\omega'_3}{ \bn'\!\cdot\! p} \,.\nn
\eea
The variable $z$ corresponds to the fraction of the $Q\bar Q$ pair  light-cone momentum that  $H$ carries away. The variables $u$ and $v$ correspond to the fraction of the total $Q\bar Q$ light-cone momentum carried by each of the heavy quarks in the $Q\bar Q$ pair. These variables do not have to be the same. The only constraint on the momentum is that the difference of the total light-cone momentum of the two heavy quark--anti-quark pairs is zero. The expression above can be inverted:
\bea
\label{invQQfragfun}
D_{a\{ 1,8\}}^{Q\bar Q}(u,v,z)&=& \frac{1}{8} \int d\omega'_1d\omega'_2d\omega'_3 d\omega'_4
\delta(\omega'_1-\omega'_2-\frac{ \bn'\!\cdot\! p}{z})\delta(\omega'_2-\frac{ \bn'\!\cdot\! p}{z}(v-1))\,\delta(\omega'_4-\frac{ \bn'\!\cdot\! p}{z}u)\nn \\
&&\times
 \, \langle 0| 
\bar \chi_{n^\prime, \omega'_2}   \Gamma^{a (\nu) } \{ \mathbbm{1},T^A\} \chi_{n^\prime,  \omega'_1}
{\cal P}^H_{n',\bn'\cdot p}
\bar \chi_{n^\prime,  \omega'_4}   \Gamma^a_{ (\nu) } \{ \mathbbm{1},T^A\}  \chi_{n^\prime,  \omega'_3} |0\rangle \,.
\eea
This definition of the DPFF is proportional to the one in Ref.~\cite{KQS} with the following variable redefinition: $u \to (1+\zeta)/2$ and $v \to (1+\zeta')/2$.
Substituting  Eqs.~(\ref{quarkpdf}, \ref{QQfragfun}) into Eq.~(\ref{fragcontQQqqfierzex}) gives a generalized factored form for the SD production cross section
\bea
\label{DPFfact}
\frac{d^2\sigma}{dy d p^2_\perp} &=& \frac{1}{2}\int d x_1 d x_2  \frac{d z}{z} du dv \bigg(\frac{\bn'\!\cdot\! p}{z}\bigg)^3 
\frac{1}{(4 \pi)^2}\hat \sigma^{a\{1,8\}}(x_1, x_2, z, u, v, p_\perp,y)\\
&& \times  f_{q/p}(x_1) f_{\bar{q}/p}(x_2) D^{Q \bar Q}_{a\{1,8\}}(u,v,z) \,,\nn
\eea
where the factor of $(\bn'\!\cdot\! p/z)^3$ comes from switching from the dimension-full variables $\omega_i$, $\bar \omega_i$ and $\omega'_i$ to dimensionless variables $x_1$, $x_2$, $z$, $u$, $v$. Our result agrees with the factorization formula in Ref.~\cite{KQS}.

\section{Evolution Equations for the DPFF}

\begin{figure}
\begin{center}
\includegraphics[width=1.5in]{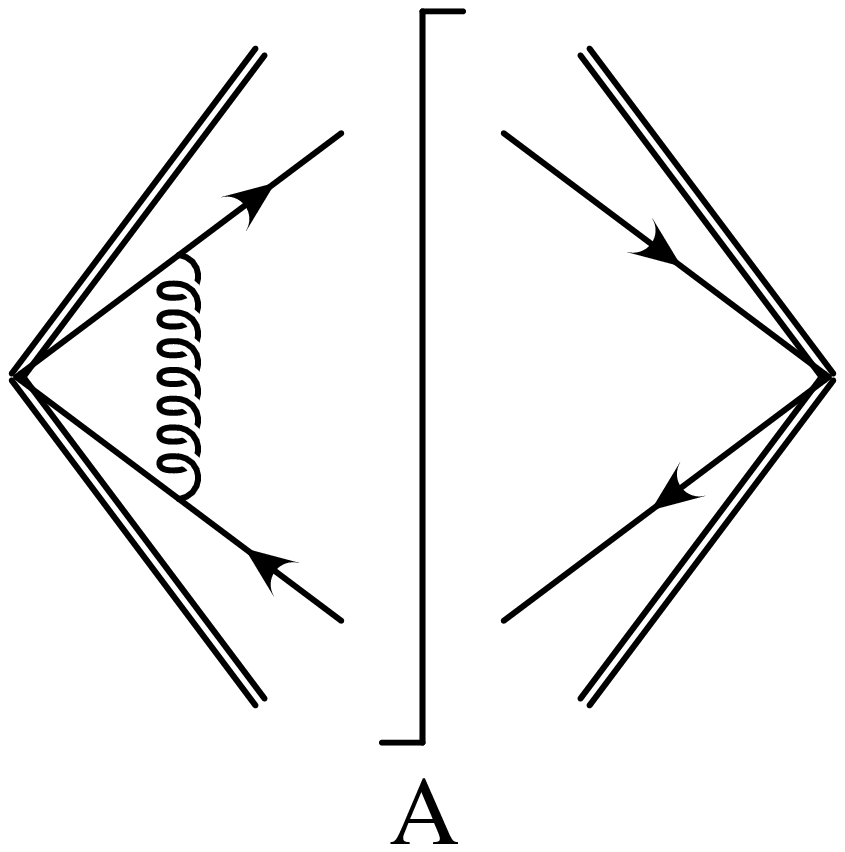}\quad\includegraphics[width=1.5in]{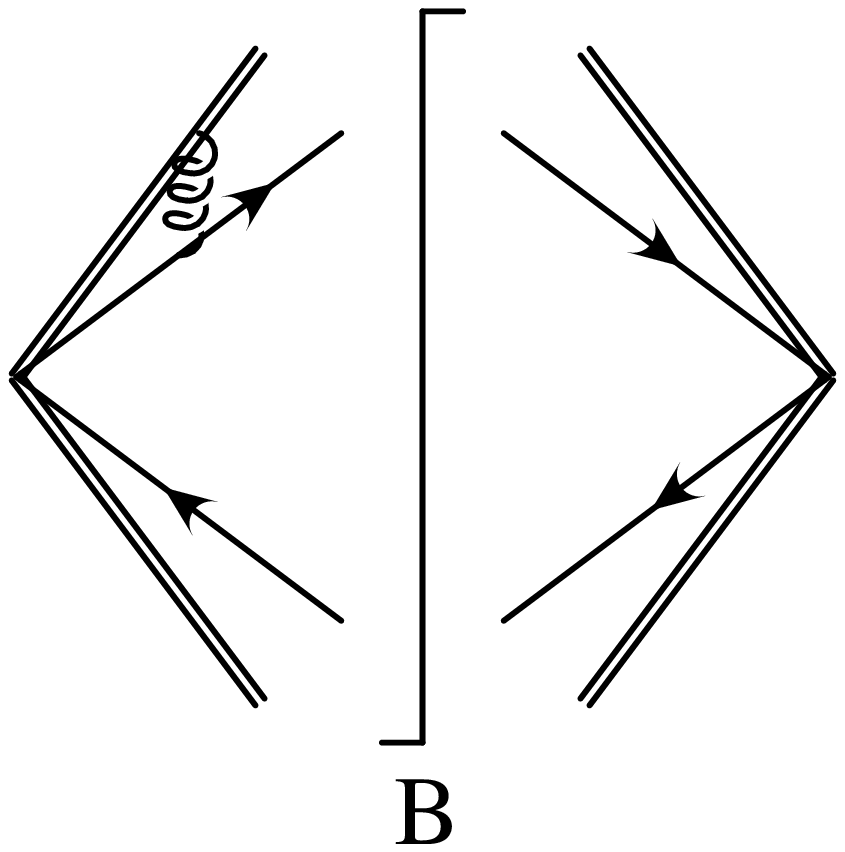}\quad\includegraphics[width=1.5in]{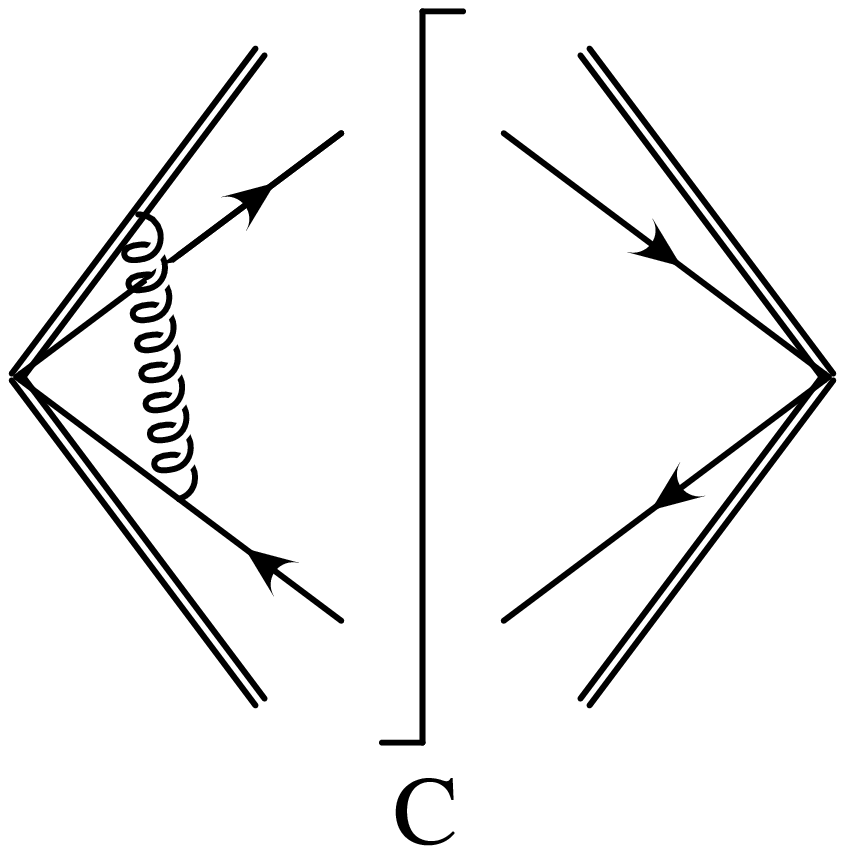}\quad\includegraphics[width=1.5in]{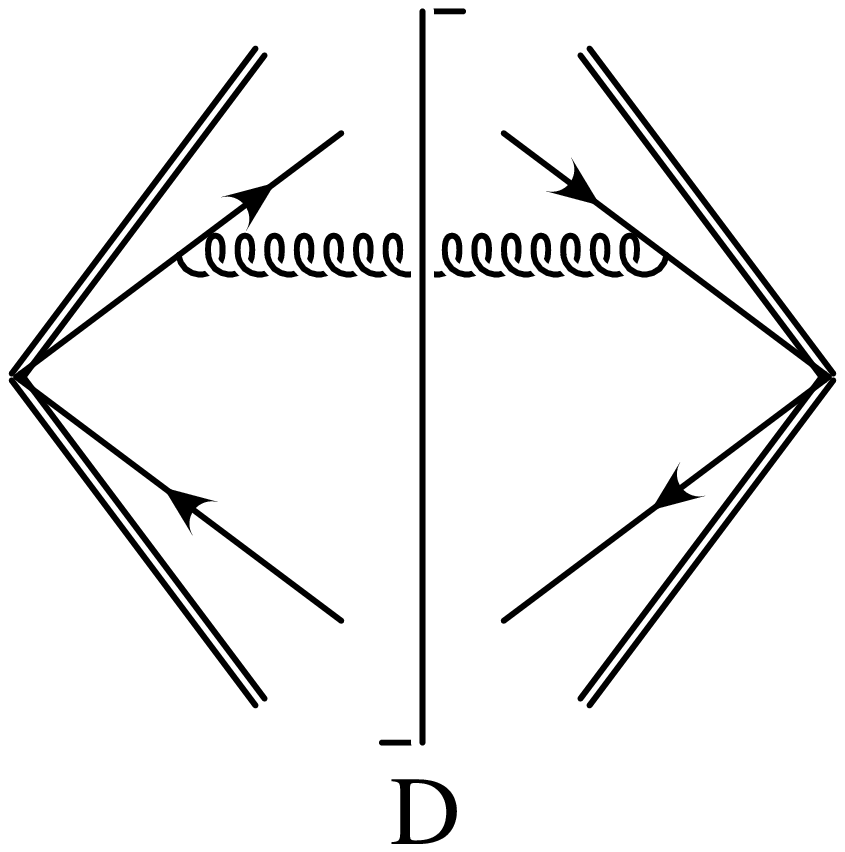}\\
\includegraphics[width=1.5in]{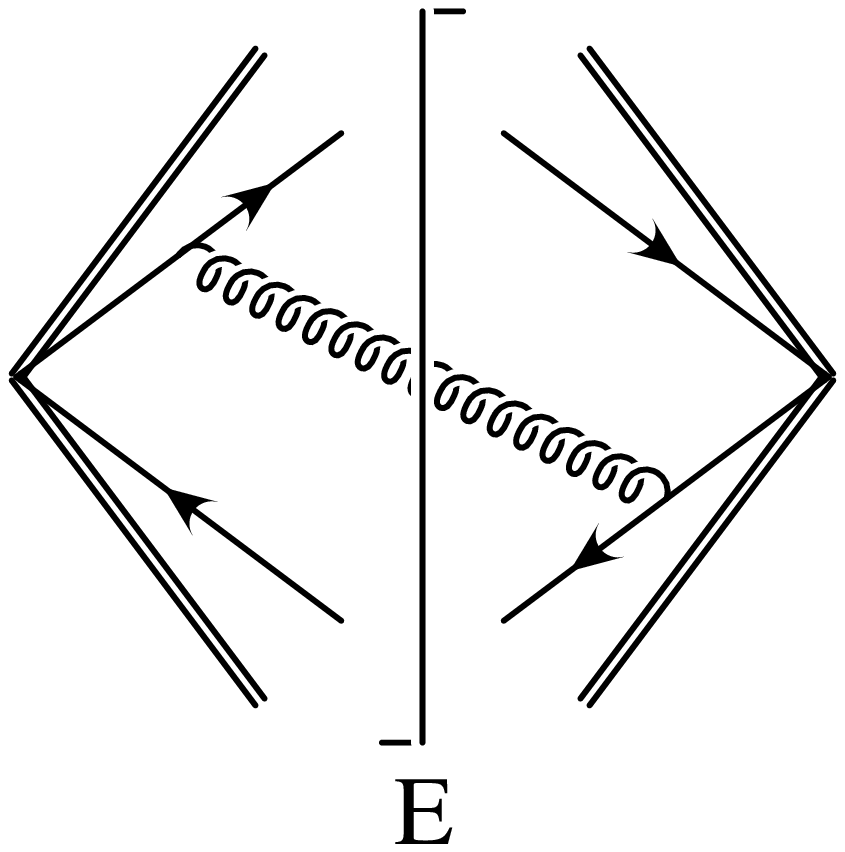}\quad\includegraphics[width=1.5in]{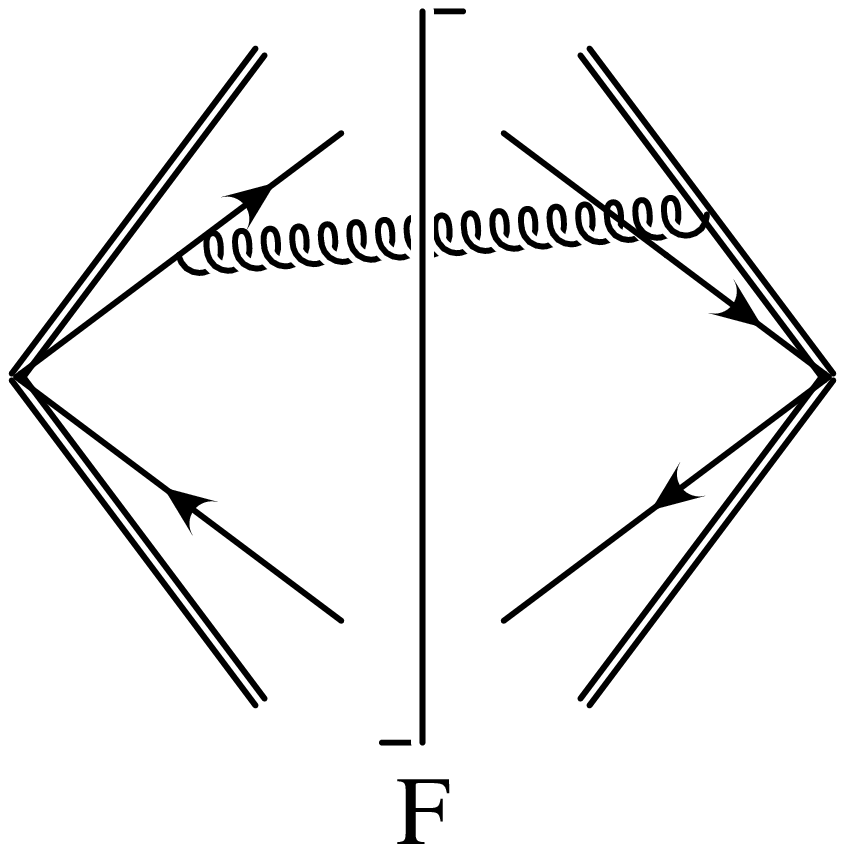}\quad\includegraphics[width=1.5in]{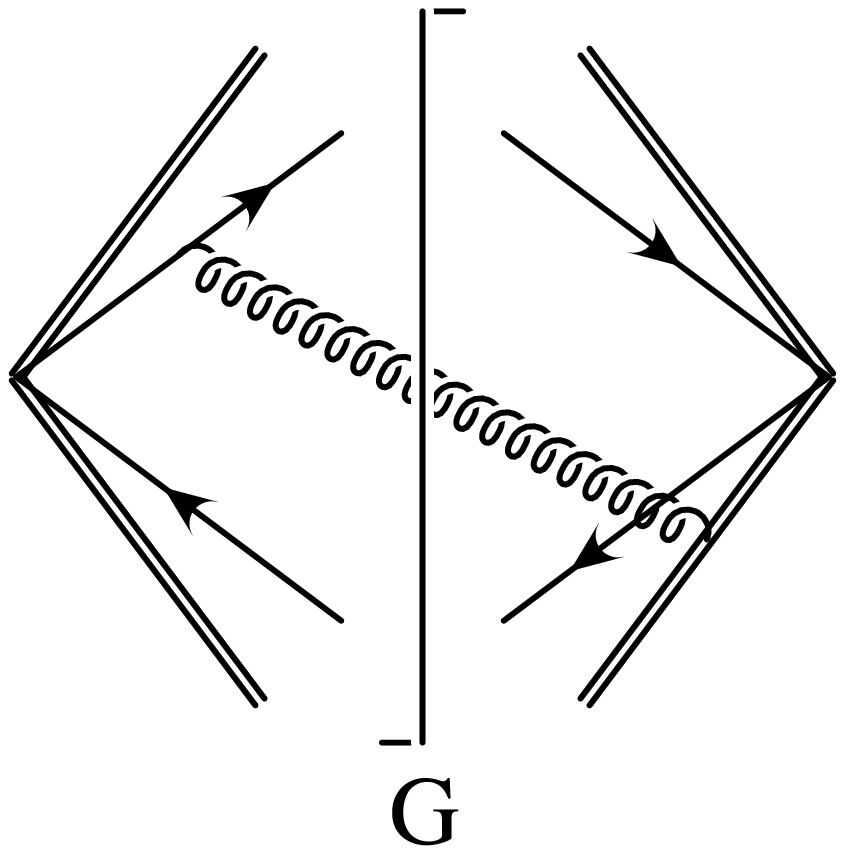}
\caption{The diagrams we need for the one-loop running. Not shown are the diagrams which are mirror images with respect to
the horizontal and vertical axes. Diagram A reflected about a horizontal (vertical) is denoted in the text by $\hat A (\bar A)$.  }
\label{diagrams}
\end{center}
\end{figure}

In this section, we derive the evolution equations for the $D^{Q \bar Q}_{3\{1,8\}}(u,v,z)$ DPFFs. We consider these DPFFs in particular because they are most 
relevant to color-singlet ${}^3S_1$ production.  The diagrams for computing the one-loop anomalous dimensions are shown in 
Fig.~\ref{diagrams}A-G. In these diagrams the single lines represent SCET collinear fields and the double lines are Wilson lines. In addition to the seven diagrams,  it is possible to generate additional diagrams by reflecting a diagram about the horizontal or vertical axes, or both. We refer to diagrams obtained from those in Fig.~\ref{diagrams} by reflecting about the horizontal axis by adding a hat, 
e.g. $\hat B$, diagrams obtained by reflecting about the vertical axis by adding a bar, e.g. $\bar B$, and by doing both reflections by adding a hat and bar, e.g. $\bar {\hat B}$. Note that $A = \hat A$, $D = \bar D$, and $\bar E =\hat E$ and these do not constitute distinct diagrams. The remaining diagrams have distinct images under the three possible reflections. As mentioned above we focus on those operators which have the Dirac structure $\bnslash'\gamma_\nu^\perp$, since these are most relevant to the production of $Q\bar Q$ pairs in $^3S_1$ configurations. Note the this Dirac structure does not mix with other Dirac structures due to the symmetries of SCET. However, we allow for both octet and singlet operators since they do mix.

\subsection{Virtual Diagrams}

The divergent pieces of diagrams $A$ and $\bar A$ vanish at leading power. If we consider the color-singlet operator and calculate diagrams $B$ and $C$ only, the IR divergences cancel and the UV divergence would lead to an anomalous dimension identical to the ERBEL~\cite{BL,ER} evolution for light-cone wavefunctions. 
In general the infrared divergences in diagrams $B$ and $C$ do not cancel because the $C$ diagram has a different color factor for the color-octet operator. Instead the IR divergences cancel between a number of different diagrams in a non-trivial way.  

In the diagrams in Fig.~\ref{diagrams} the quark and anti-quark on the left-hand side of the cut have {\it outgoing} momentum $p^\mu_4$ and $p^\mu_3$ respectively, while the quark and anti-quark on the right-hand side of the cut have {\it incoming} momentum $p^\mu_1$ and $p^\mu_2$ respectively. We  express the large components of these momenta in terms of momentum fractions:
\bea
x &=& \frac{P}{\bn'\cdot (p_1+p_2)} \\
\lambda  &=& x \frac{\bn'\cdot p_4}{P} \nn \\
\xi &=& x \frac{\bn' \cdot p_1}{P} \,. \nn
\eea
where $P$ is the large light-cone momentum component of the final state $Q\bar{Q}$ pair.
Another technical complication is that individual diagrams have rapidity divergences that cancel in the sum over diagrams, but must be regulated at intermediate stages of the calculation. This is accomplished here by adopting the rapidity regulator of Refs.~\cite{Chiu:2011qc,Chiu:2012ir}

The result of evaluating diagram $B$ is  
\bea
M_B&=& \frac{\alpha_s}{2\pi} \frac{C_F}{\epsilon}\bigg[ \frac{1}{\eta}+\ln\bigg(\frac{z \nu}{u P}\bigg)+1\bigg] \bigg(\frac{z}{2P}\bigg)^3 \delta(1-z/x)\delta(\lambda-u)\delta(\xi-v) O^{j} \, ,
\eea
where $j=1,8$ indicates color-structure, and 
\beq
O^{(1,8)} = \bar \xi_{n'} \frac{\bnslash' \gamma^\nu_{\perp'}}{2}\{1,T^a\} \xi_{n'}
 \bar \xi_{n'} \frac{\bnslash' \gamma_\nu^{\perp'}}{2}\{1,T^a\} \xi_{n'}
\eeq
is a combination of SCET spinors in either a color-singlet or color-octet combination.
The symmetric diagrams can be found via the simple replacements:
\bea\label{subs}
M_{\bar B} &=& M_B(u\leftrightarrow v, \lambda \leftrightarrow \xi), \\
M_{\hat B} &=& M_B(u\leftrightarrow \bar u, \lambda \leftrightarrow \bar\lambda), \nn \\
M_{\bar{\hat B}} &=& M_B(u\leftrightarrow \bar v, \lambda \leftrightarrow \bar\xi), \nn
\eea
where $\bar u = 1-u$, $\bar v = 1-v$, $\bar \lambda = 1-\lambda$, and $\bar \xi = 1- \xi$.
We find that the sum is
\bea
M_B+M_{\bar B}+M_{\hat B}+M_{\bar{\hat B}} &=& \\
&& \hspace{-12ex} 
\frac{\alpha_s}{2\pi} \frac{C_F}{\epsilon}
\bigg[ \frac{4}{\eta}+\ln\bigg(\frac{z^4 \nu^4}{u\bar u v \bar v P^4}\bigg)+4\bigg] \bigg(\frac{z}{2P}\bigg)^3 \delta(1-z/x)\delta(\lambda-u)\delta(\xi-v) O^{j}  \, .\nn
\eea

Now we consider diagram  $C$ and its reflections. These diagrams have different color factors depending upon whether or not
the operator is color-singlet or color-octet. 
We denote the color factors by $\beta^{(1,8)}$, where $\beta^{(1)}=C_F$ and $\beta^{(8)}=-\frac{1}{2N_c} $. Diagram $C$ yields
\bea
M_{C}&=& -\frac{\alpha_s}{2\pi} \frac{\beta^{(j)}}{\epsilon}\bigg\{\bigg[ \frac{1}{\eta}+\ln\bigg(\frac{z\nu}{\bar u P}\bigg)\bigg]\delta(\lambda-u)-\frac{\bar u}{\bar \lambda}\frac{\theta(u-\lambda)}{(u-\lambda)_+}\bigg\} \bigg(\frac{z}{2P}\bigg)^3 
\delta(v-\xi)\delta(1-\frac{ z}{x}) O^{j}\, . \nn \\
&&
\eea
Again the diagrams related by symmetry can be obtained by making the  substitutions in Eq.~(\ref{subs}).
Note that diagrams $B$ and $C$ and their reflections do not lead to any mixing between singlet and octet operators. 

As before, the individual diagrams are not IR finite, and the result of summing diagram $C$ and its reflections is
\bea
M_C+M_{\bar C}+M_{\hat C}+M_{\bar{\hat C}}&=& -\frac{\alpha_s}{2\pi} \frac{\beta^{(j)}}{\epsilon}
\bigg\{\bigg[ \frac{4}{\eta}+\ln\bigg(\frac{z^4 \nu}{u\bar u v \bar v P^4}\bigg)\bigg]\delta(\lambda-u)\delta(\xi-v)-\bigg[\frac{u}{\lambda}\frac{\theta(\lambda-u)}{(\lambda-u)_+}\nn  \\
&& \hspace{-24ex}+\frac{\bar u}{\bar \lambda}\frac{\theta(u- \lambda)}{(\bar \lambda-\bar u)_+}\bigg]\delta(\xi-v)
- \bigg[\frac{v}{\xi}\frac{\theta(\xi-v)}{(\xi-v)_+}
+\frac{\bar v}{\bar \xi}\frac{\theta(v-\xi)}{(\bar \xi-\bar v)_+}\bigg]  \delta(\lambda-u)\bigg\} 
 \bigg(\frac{z}{2P}\bigg)^3 
 \delta(1-z/x) O^{j}  \, . 
\eea
Notice that for the color-singlet operator, but not the color-octet operator, the rapidity divergences (i.e., the $1/\eta$ poles) and   the corresponding logarithmic terms cancel between diagrams $B$ and $C$ and their reflections. The remaining terms lead to an evolution equation kernel that is similar to that of a light-cone wave function~\cite{BL,ER}. For the color-octet operator the rapidity divergences in diagrams $B$ and $C$ cancel against the rapidity divergences in the real emission graphs which we turn to now.
 
\subsection{Real Radiation}
Now consider the real radiation coming from diagrams $D-G$ and their reflections.
We introduce the color factor matrices:
\[
\beta_{ij}=\left(
\begin{array}{cc}
 \beta_{11}  & \beta_{18}     \\
  \beta_{81}& \beta_{88}    \\  
\end{array}
\right)
=\left(
\begin{array}{cc}
 0  &  \frac{C_F}{2N_c}     \\
1 & \frac{N_c^2-2}{2N_c}     \\  
\end{array}
\right)
,\qquad
\bar \beta_{ij}=\left(
\begin{array}{cc}
 0  &    \frac{C_F}{2N_c}   \\
 1  & -\frac{1}{N_c}     \\  
\end{array}
\right) \, .
\]
The first index, $i=1$ or $8$,  refers to the color state of the initial and final  state quarks in the diagram and the second index, $j$, refers to the color-structure of the operator. We present only the diagrams shown in Fig.~\ref{diagrams}: 
\bea
M_D&=& \frac{\alpha_s}{2\pi} \left( \frac{z}{2P}\right)^3 \frac{1}{\epsilon_{UV}} \beta_{ij} \frac{x^2}{z^2}\frac{1-z/x}{\lambda \xi} \delta\left(\bar v-\frac{z}{x}\bar \xi\right) \delta\left(\bar u -\frac{z}{x}\bar \lambda\right)O^{j},
\\\nonumber\\
M_E&=& -\frac{\alpha_s}{2\pi} \left( \frac{z}{2P}\right)^3 \frac{1}{\epsilon_{UV}}\bar  \beta_{ij} \frac{x^2}{z^2}\frac{1-z/x}{\lambda \bar \xi} \delta\left(v-\frac{z}{x} \xi \right) \delta\left(\bar u -\frac{z}{x}\bar \lambda\right)O^j,
\nonumber\\\nonumber\\
M_F &=& \frac{\alpha_s}{2\pi} \left( \frac{z}{2P}\right)^3 \frac{1}{\epsilon_{UV}} \beta_{ij}\bigg\{ -\bigg[ \frac{1}{\eta}+\ln\bigg(\frac{z \nu}{uP}\bigg)\bigg]\delta(1-z/x)
+\frac{u x}{\lambda z}\frac{\theta(1-z/x)}{ (1-z/x)_+} \bigg\}\nn\\
&& \hspace{25ex}\times \delta\left(\bar v-\frac{z}{x}\bar \xi\right) \delta\left(\bar u -\frac{z}{x}\bar \lambda\right)O^j,
\nonumber\\\nonumber\\
M_G&=&- \frac{\alpha_s}{2\pi} \left( \frac{z}{2P}\right)^3 \frac{1}{\epsilon_{UV}}\bar  \beta_{ij} \bigg\{ -\bigg[ \frac{1}{\eta}+\ln\bigg(\frac{z\nu}{u P}\bigg)\bigg]\delta(1-z/x)
+\frac{u x}{\lambda z}\frac{\theta(1-z/x)}{ (1-z/x)_+}\bigg\}\nn\\
&& \hspace{25ex}\times \delta\left(v-\frac{z}{x}\xi\right) \delta\left(\bar u -\frac{z}{x}\bar \lambda\right)O^j
\, . \nn
\eea
The reflections of these diagrams can be determined by
the same substitutions given in Eq.~(\ref{subs}):
As previously mentioned, the rapidity divergences cancel in the sum of
the collinear diagrams. This is as it must be since the process we are considering is bereft of any soft 
sector which usually supplies the mechanism for the cancellation of rapidity divergences.

\section{Renormalization}
The DPFF is renormalized multiplicatively as follows 
\beq
D_i^0(u,v,z)= \int du^\prime dv^\prime \frac{dz^\prime}{z^\prime} Z_{ij}(u,u^\prime,v,v^\prime,z/z^\prime,\mu)D_j^R(u^\prime,v^\prime,z^\prime,\mu).
\eeq
The renormalized distribution thus obeys 
\beq
\mu \frac{d}{d\mu} D_{i}^R(u^{\prime \prime},v^{\prime \prime},z^{\prime \prime},\mu)= - \int 
du^\prime dv^\prime \frac{dz^\prime}{z^\prime}\gamma_{ij} (u^{\prime \prime},u^\prime,v^{\prime \prime},v^\prime,z^{\prime \prime}/z^\prime,\mu)D_{j}^R(u^{\prime},v^{\prime},z^{\prime},\mu)
\eeq
where the  anomalous dimension is given by
\bea
\gamma_{ij}(u,u^\prime,v,v^\prime,z/z^\prime,\mu)&=& \\
&& \hspace{-10ex}\int du^{\prime \prime}dv^{\prime \prime}\frac{dz^{\prime \prime}}{z^{\prime \prime}} Z_{ia}^{-1}(u,u^{\prime \prime},v,v^{\prime \prime},z/z^{\prime \prime},\mu)\mu\frac{d}{d\mu}
Z_{aj}(u^{\prime \prime},u^\prime,v^{\prime \prime},v^\prime,z^{\prime \prime}/z^\prime,\mu) \, ,\nn
\eea
and the indices $i$ and $j$ label the singlet $(i,j=1)$ and octet $(i,j=8)$ operators.
The tree level matrix element of the DPFF using partonic states with momenta labelled by $(x,\lambda,\xi)$ is given by
\bea
D_j(u,v,z)=\left( \frac{z}{2P}\right)^3 \delta(1-z/x)\delta(\lambda-u)\delta(\xi-v)O^{j} \,.
\eea
Given the results of the previous section and 
the wave function renormalization (which in SCET is identical to QCD), 
\beq
Z_\xi=1- \frac{\alpha_s C_F}{4 \pi \epsilon},
\eeq
we find the anomalous dimensions are given by 
\bea
\gamma_{11}&=&-3\frac{\alpha_s C_F}{\pi} \delta(u-u^\prime)\delta(v-v^\prime) \delta(1-z/z^\prime) \\
&&-\frac{\alpha_s C_F}{\pi}\delta(v-v^\prime)\delta(1-\frac{ z}{z^\prime})\left[ \theta(u^\prime-u) \frac{u}{u^\prime} \frac{1}{(u^\prime-u)_+} + \theta(u-u^\prime) \frac{\bar u}{\bar u^\prime} \frac{1}{(u-u^\prime)_+} \right]\nn \\
&&-\frac{\alpha_s C_F}{\pi}\delta(u-u^\prime)\delta(1-\frac{ z}{z^\prime}) \left[\theta(v^\prime-v) \frac{v}{v^\prime}\frac{1}{(v^\prime-v)_+} + \theta(v-v^\prime) \frac{\bar v}{\bar v^\prime}\frac{1}{(v-v^\prime)_+} \right], \nn\\
&&\nn\\
\gamma_{81} &=& - \frac{\alpha_s}{\pi} \theta(1-z/z^\prime) \left(\frac{z}{z'}\right)^2 \left\{\left[  \frac{u v' +v u'}{ (1-z/z')_+} + \frac{1-z/z'}{z/z'}\right] \frac1{u'v'} \delta(\bar v-\frac{z}{z'}\bar v') \delta(\bar u -\frac{z}{z'}\bar u')\right.\nn  \\
&& \phantom{- \frac{\alpha_s}{\pi}  \frac{z'}{z}} +\left[    \frac{\bar u\bar v' +\bar v\bar u' }{(1-z/z')_+}     + \frac{1-z/z'}{z/z'}\right]  \frac{1}{\bar u'\bar v'} \delta(v-\frac{z}{z'}v') \delta(u -\frac{z }{z'}u') \nn\\
&& \phantom{- \frac{\alpha_s}{\pi}  \frac{z'}{z}}  -\left[   \frac{u\bar v' +\bar v u'   }{(1-z/z')_+}  + \frac{1-z/z'}{z/z'} \right] \frac1{u'\bar v'} \delta(v-\frac{z}{z'} v' ) \delta(\bar u -\frac{z}{z'}\bar u')\nn\\
&&\left.\phantom{- \frac{\alpha_s}{\pi}  \frac{z'}{z}} - \left[  \frac{\bar u v' + v \bar u' }{(1-z/z')_+} + \frac{1-z/z'}{z/z'} \right] \frac1{\bar u' v'} \delta(\bar v-\frac{z  }{z'}\bar v') \delta(u -\frac{z }{z'}u')\right\}.
\nn
\eea
We also have 
\beq
\gamma_{18} = \frac{C_F}{2N_c} \gamma_{81} \, .
\eeq
Finally, 
\bea
\gamma_{88} &=&  -3 \frac{\alpha_sC_F}{\pi} \delta(u-u^\prime)\delta(v-v^\prime) \delta(1-z/z^\prime)\\
&&\hspace{-1ex}+\frac{\alpha}{\pi}\frac{1}{2N_c}\delta(v-v^\prime)\delta(1-\frac{ z}{z^\prime})\left[ \theta(u^\prime-u) \frac{u}{u^\prime} \frac{1}{(u^\prime-u)_+} + \theta(u-u^\prime) \frac{\bar u}{\bar u^\prime} \frac{1}{(u-u^\prime)_+} \right]\nn \\
&&\hspace{-1ex}+ \frac{\alpha_s}{\pi}\frac{1}{2N_c}\delta(u-u^\prime)\delta(1-\frac{ z}{z^\prime}) \left[\theta(v^\prime-v) \frac{v}{v^\prime}\frac{1}{(v^\prime-v)_+} + \theta(v-v^\prime) \frac{\bar v}{\bar v^\prime}\frac{1}{(v-v^\prime)_+} \right]\nn\\
&&\hspace{-1ex} - \frac{\alpha_s}{\pi} \theta(1-z/z^\prime) \left(\frac{z}{z'}\right)^2 \left\{\frac{N_c^2 - 2}{2 N_c} \left[  \frac{ u v' +v u'}{ (1-z/z')_+} + \frac{1-z/z'}{z/z'}  \right] \frac1{u'v'} \delta(\bar v-\frac{z}{z'}\bar v') \delta(\bar u -\frac{z}{z'}\bar u')\right.  \nn\\
&&\hspace{-1ex} \phantom{-\frac{\alpha_s}{\pi} \frac{z'}{z} }  +\frac{N_c^2 - 2}{2 N_c}\left[   \frac{ \bar u\bar v' +\bar v\bar u'}{(1-z/z')_+}     + \frac{1-z/z'}{z/z'}  \right]  \frac{1}{\bar u'\bar v'} \delta(v-\frac{z}{z'}v') \delta(u -\frac{z }{z'}u') \nn\\
&&\hspace{-1ex} \phantom{-\frac{\alpha_s}{\pi} \frac{z'}{z} }  +\frac1{N_c}\left[    \frac{u\bar v'  +\bar v u'  }{(1-z/z')_+}  + \frac{1-z/z'}{z/z'}   \right] \frac1{u'\bar v'} \delta(v-\frac{z}{z'} \xi ) \delta(\bar u -\frac{z}{z'}\bar u')\nn\\
&&\hspace{-1ex}\left. \phantom{-\frac{\alpha_s}{\pi} \frac{z'}{z} } +\frac1{N_c} \left[ \frac{\bar u v' + v \bar u' }{(1-z/z')_+} + \frac{1-z/z'}{z/z'}\right] \frac1{\bar u' v'} \delta(\bar v-\frac{z  }{z'}\bar v') \delta(u -\frac{z }{z'}u')\right\}. \nn
\eea

The anomalous dimensions computed in this section are new and are one of our main results. Evolving these equations from the scale $p_\perp$ to $m_Q$ will allow us to resum logs of $p_\perp/m_Q$, but clearly solving these differential equations will be a complicated task that is beyond the scope of this paper. After evolving down to the scale $m_Q$, one must match the DPFF onto NRQCD matrix elements, which is the subject of the next section.

\section{Matching onto NRQCD}

At the scale $2 m_Q$ the heavy quark mass is integrated out by matching $D^{Q\bar Q}_{i}(u,v,z)$ defined in \SCETm onto NRQCD. The SCET fields in the definition of this function contain light-like Wilson lines, which the NRQCD operators inherit. In the case where the NRQCD operators are in a color-singlet configuration the Wilson lines cancel and we arrive at a standard NRQCD long-distance matrix element. However, in the case where the NRQCD operators are in a color-octet configuration, the Wilson lines do not cancel. The presence of the Wilson lines in the color-octet operator ensures the proper infrared behavior of the operator matrix element~\cite{Nayak:2005rw,Nayak:2005rt}. However, because of the presence of the Wilson lines, these color-octet production operators are different from the ones introduced in Ref.~\cite{Bodwin:1994jh}.

To perform the matching of $D^{Q \bar Q}_{i}(u,v,z)$ onto NRQCD we adapt the framework for treating heavy quark effective theory in a boosted frame~\cite{Fleming:2007qr,Fleming:2007xt} to NRQCD in a boosted frame. Consider first a heavy quark field in NRQCD.  In the heavy quark rest frame its four-velocity is 
\begin{equation}
v^\mu = (1,0,0,0)= \frac{1}{2}n'^\mu + \frac{1}{2}\bn'^\mu\,.
\end{equation}
The heavy quark momentum can be expressed as 
\begin{equation}
p^\mu = m_Qv^\mu + \tilde k^\mu + k^\mu,
\end{equation} 
where $v^\mu$ is the four-velocity of the heavy quark with $v^2=1$. Using the formalism of Ref.~\cite{Luke:1999kz}, the large components of the momentum become labels, denoted above by $m_Qv^\mu + \tilde k^\mu$, leaving derivatives acting on the field to scale as $k^\mu\sim m_Q \beta^2$, where the relative speed of the heavy quark and anti-quark (not to be confused with the heavy quark four-velocity $v^\mu$)  is $\beta\ll 1$.  NRQCD gluons and massless quarks have momenta that  scales as $m_Q \beta^2$ (i.e., they do not have labels).

In a frame in which the heavy quark--anti-quark pair is boosted in the direction $n'$ by a large factor $Q$, we have
\begin{equation}
\label{velocity}
v^\mu = \frac{1}{4} \frac{Q}{m_Q} n'^\mu +\frac{m_Q}{Q} \bn'^\mu\,,
\end{equation}
with a similar boosting for the other pieces of the momenta, such as
\begin{equation}
k^\mu_{boost} =  \frac{1}{4} \frac{Q}{m_Q}\bn'\!\cdot\! k_{rest} \, n'^\mu + \frac{m_Q}{Q}n'\!\cdot\! k_{rest} \, \bn'^\mu+k_{rest}^{\perp'\mu}\,,
\end{equation}
where $k_{rest}$ are the components of momentum in the rest frame.  So, for example, the NRQCD residual momentum scales as
\beq
k^\mu_{boost} \sim \left(Q\beta^2, \frac{m_Q^2\beta^2}Q, m_Q\beta^2\right).
\eeq
To match onto NRQCD we need to identify which components of the massive SCET momenta match onto components of the labels and residual momentum in NRQCD.
A generic massive SCET momentum in the $n'$ direction is $p^\mu = \tilde p^\mu + r^\mu$, where the label momentum is $\tilde p^\mu = \bn'\!\cdot\!\tilde p \, n'^{\mu} /2 + \tilde p_{\perp'}^\mu$ with scaling $ \bn'\!\cdot\!\tilde p  \sim Q$, and $ \tilde p_{\perp'}^\mu \sim m_Q \sim Q \lambda$. The residual momentum scales as $r^\mu \sim m_Q^2/Q \sim Q \lambda^2$. Working in a frame where $ p_{\perp'}^\mu=0$, the label momentum $\bn'\!\cdot\!\tilde p$ in massive SCET matches onto the large component of the boosted velocity $m_Q\, \bn'\!\cdot\! v$ in NRQCD.
The SCET residual momentum will be split into label and residual NRQCD components.  

Next we consider how bilinears of massive SCET fields match onto bilinears in NRQCD. The generic SCET bilinear can be written as
\beq
\label{bilinearmatch}
 \bCH{n'}{\omega_2} \Gamma^{i(\nu)} \{ \mathbbm{1},T^A\}   \CH{n'}{\omega_1}
= \bar \xi_{n',\tilde p_2}W_{n'}\delta(\bn'\!\cdot\! \cP^\dagger-\omega_2) \Gamma^{i(\nu)}  \{ \mathbbm{1},T^A\} 
\delta(\bn'\!\cdot\! \cP-\omega_1) W_{n'}^\dagger \xi_{n',\tilde p_1}.
\eeq
When the virtuality drops below $m^2$ collinear parton splitting is no longer possible. However, the collinear Wilson
lines in \SCETm match onto another set of Wilson lines which can be thought of as arising from boosting
ultra-soft Wilson lines that arise in NRQCD after performing a BPS field redefinition.
We call such gluons ultra-collinear and designate the field as $A^\mu_{uc}(x)$. Note that the momentum scaling of these gluons is down by a factor of $\beta^2$ compared to the usual collinear SCET gluons. 
This is because in matching we have reduced the virtuality of the external states from $m^2$ to $m^2\beta^4$.
  As a result we find that the SCET Wilson line matches onto an NRQCD Wilson line
\begin{equation}
W_{n'}(x) \to W_{uc}(x)= \textrm{P} \exp\bigg[ ig \int^x_{-\infty} ds \, \bn'\!\cdot\! A_{uc}(\bn' s)\bigg]\,.
\end{equation} 
In the matching the delta functions in Eq.~(\ref{bilinearmatch}) fix $\omega$: 
\bea
\delta(\bn'\!\cdot\! \cP-\omega) W_{n'}^\dagger(x) \xi_{n',\tilde p}(x)& \to & \delta(\bn'\!\cdot\! \cP-\omega) W^\dagger_{uc}(x) \xi_{n',\tilde p}(x) \nn \\
&=&  W^\dagger_{uc}(x)\delta(\bn'\!\cdot\! \cP-\omega) \xi_{n',\tilde p}(x)\nn \\
&=&  \delta(m_Q \bn'\!\cdot\! v-\omega) W^\dagger_{uc}(x) \xi_{n',\tilde p}(x) \,.
\eea
We were able to push the delta function past the ultra-collinear Wilson line since $\bn'\!\cdot\! \cP$ acting on the Wilson line is again down by $\beta^2$ compared to when it acts on the quark field.

Finally we consider the matching of the bilinears of  quark and anti-quark fields. For this step we can ignore the Wilson lines and delta functions, and focus on
\begin{equation}
\label{relation}
 \bar \xi_{n',\tilde p_2} \Gamma^{i(\nu)}  \{ \mathbbm{1},T^A\}  \xi_{n',\tilde p_1}\,.
\end{equation}
We reintroduce the large phase and relate the SCET fields to standard 4-component QCD fields~\cite{Rothstein:2003wh}:
\begin{equation}
e^{-i \tilde p \cdot x} \xi_{n',\tilde p}(x) = \bigg(1-\frac{i\dslash_\perp + m_Q}{i\bn'\cdot \partial}\frac{\bnslash '}{2}\bigg) \psi(x).
\end{equation}
Inserting this identity into Eq.~(\ref{relation}) and simplifying we find 
\begin{equation}
e^{-i (\tilde p_1- \tilde p_2) \cdot x}\bar \xi_{n',\tilde p_2} \Gamma^{i(\nu)}  \{ \mathbbm{1},T^A\}  \xi_{n',\tilde p_1}
= \bar\psi(x) \Gamma^{i(\nu)}  \{ \mathbbm{1},T^A\} \psi(x) \,.
\end{equation}
Matrix elements of the bilinear in terms of QCD fields can then be related to matrix elements of bilinears of NRQCD fields using the techniques of Ref.~\cite{Bratten:1996he}, with the matching schematically given by
\begin{equation}
 \bar\psi(x) \Gamma^{i(\nu)}  \{ \mathbbm{1},T^A\} \psi(x) \to {\cal C}_i^{(\nu)} \chi_v^\dagger \Sigma(\vec \sigma,\vec{D})\{ \mathbbm{1},T^A\}\psi_v  + \textrm{h.c.}\,,
\end{equation}
where ${\cal C}_i^{(\nu)}$ are matching coefficients, $\psi_v$ and $\chi_v$ are the heavy quark and anti-quark fields respectively, $\vec \sigma$ are the Pauli matrices, $\vec{D}$ is the covariant derivative in NRQCD, and $\Sigma(\vec \sigma,\vec{D})$ is a fixed function of $\vec \sigma$ and $\vec{D}$ at a given order in the NRQCD expansion. We have suppressed vector indices on ${\cal C}_i^{(\nu)}$ and $\Sigma(\vec \sigma,\vec{D})$.
Putting all these pieces together, the matching from SCET onto NRQCD has the form
\bea
&& \bCH{n'}{\omega_2} \Gamma^{i(\nu)} \{ \mathbbm{1},T^A\}   \CH{n'}{\omega_1}\\
&&\hspace{10ex} \to {\cal C}_i^{(\nu)}  \delta(m_Q \bn'\!\cdot\! v-\omega_1) \delta(m_Q \bn'\!\cdot\! v+\omega_2)\chi_v^\dagger W_{uc} \Sigma(\vec \sigma,\vec{D})\{ \mathbbm{1},T^A\}W^\dagger_{uc} \psi_v  + \textrm{h.c.}\nn 
\eea
For a color-singlet configuration $W_{uc}W^\dagger_{uc}=1$ and the Wilson lines cancel in NRQCD.

We now match a generic DPFF  onto NRQCD operators
\bea
\label{xxxx}
D^{QQ}_{(1,8)}(u,v,z) &  &  \\
& &\hspace{-15ex} \to {\cal D}_{i} z \, \delta\left(2m_Q \bn'\!\cdot\! v-\frac{ 2m_Q \bn'\!\cdot\! v}{z}\right)\delta\left(m_Q \bn'\!\cdot\! v+\frac{2m_Q \bn'\!\cdot\! v}{z}(v-1)\right)\,\delta\left(m_Q \bn'\!\cdot\! v-\frac{2m_Q \bn'\!\cdot\! v}{z}u\right)\nn \\
& & \hspace{-10ex}\times
 \, \langle 0| 
\chi^\dagger_{v} W_{uc} \Sigma(\vec \sigma,\vec{D}) \{ \mathbbm{1},T^A\} W^\dagger_{uc} \psi_{v}
{\cal P}^{H_{v}}
\psi^\dagger_{v}  W_{uc} \Sigma(\vec \sigma,\vec{D}) \{ \mathbbm{1},T^A\}  W^\dagger_{uc} \chi_{v} |0\rangle \nn \\
& &  \hspace{-15ex} = \frac{{\cal D}_i}{Q^3}  \delta(1-z) \delta\left(\frac{1}{2}-v\right)\delta\left(\frac{1}{2}-u\right)\nn \\
& & \hspace{-10ex}\times
 \, \langle 0| 
\chi^\dagger_{v} W_{uc} \Sigma(\vec \sigma,\vec{D}) \{ \mathbbm{1},T^A\} W^\dagger_{uc} \psi_{v}
{\cal P}^{H_{v}}
\psi^\dagger_{v}  W_{uc} \Sigma(\vec \sigma,\vec{D}) \{ \mathbbm{1},T^A\}  W^\dagger_{uc} \chi_{v} |0\rangle \,,  \nn
\eea
where ${\cal D}_i$  are matching coefficients, and
\begin{equation}
{\cal P}^{H_{v}}=\sum_{X_{uc}} |H_v +X_{uc} \rangle \langle H_v +X_{uc}|\,,
\end{equation}
with $H_v$ the NRQCD quarkonium state and $X_{uc}$ ultra-collinear states.
As pointed out above, if the heavy quark bilinear is in a color-singlet configuration all ultra-collinear Wilson lines cancel and we are left with the standard NRQCD production matrix elements. However, if the heavy quark bilinear is in a color-octet configuration, the ultra-collinear Wilson lines do not cancel and we obtain color-octet NRQCD matrix elements with light-like Wilson lines as first proposed in Refs.~\cite{Nayak:2005rw,Nayak:2005rt}.

As a concrete example let us consider the matching of $D^{QQ}_{1}(u,v,z)$ to the color-singlet $^3S_1$ operator at leading order in the NRQCD expansion. The Dirac structure we need to match is
\bea
&&\bar \chi_{n^\prime, \omega'_2}  \frac{\bnslash'}{2}  \gamma^\nu_\perp \chi_{n^\prime,  \omega'_1}
{\cal P}^H_{n',Q}
\bar \chi_{n^\prime,  \omega'_4}  \frac{\bnslash'}{2}  \gamma^\perp_\nu  \chi_{n^\prime,  \omega'_3}  \to \\
&&\frac{Q^2}{12 m_Q} \delta(m_Q \bn'\!\cdot\! v-\omega'_1)  \delta(m_Q \bn'\!\cdot\! v+\omega'_2)\delta(m_Q \bn'\!\cdot\! v+\omega'_3)\delta(m_Q \bn'\!\cdot\! v-\omega'_4) \chi_v^\dagger \sigma^i \psi_v   {\cal P}^{H_{v}}\psi_v^\dagger \sigma^i \chi_v
\,.\nn 
\eea 
Thus,
\bea
\label{match}
D^{QQ}_{1}(u,v,z) & \to & \frac{1}{12 Q m_Q}  \delta(1-z) \delta\left(\frac{1}{2}-v\right)\delta\left(\frac{1}{2}-u\right)
 \langle 0| 
 \chi_v^\dagger \sigma^i \psi_v   {\cal P}^{H_{v}}\psi_v^\dagger \sigma^i \chi_v
|0\rangle \,.  
\eea
If running is neglected and Eq.~(\ref{match}) is inserted into Eq.~(\ref{DPFfact}) then we recover the results of the NRQCD factorization formalism.

\section{Conclusions}
In this paper we used SCET to derive a factorization formula for fragmentation and double fragmentation
production of quarkonium. In addition, we have shown that production in the regime
where the hard scattering scale is much larger than $p_\perp$ is suppressed.   
We have presented for the first time the anomalous dimensions of the two by two system of
DPFFs that are relevant for the production of $^3S_1$ states and showed how to match the DPFF onto NRQCD production operators. The running of the DPFF sums logs of $p_\perp/m_Q$ and could have a large effect on the quarkonium production rate. A future publication will  calculate the effects of the running of the DPFF on both the production rate and
the polarization.

\acknowledgements

We thank Iain Stewart and Wouter Waalewijn for useful conversations. SF  was supported in part by the Director, Office of Science, Office of Nuclear Physics, of the U.S. Department of Energy under grant numbers DE-FG02-06ER41449 and DE-FG02-04ER41338. SF  also acknowledges support from the DFG cluster of excellence ``Origin and structure of the
universe''. AKL was supported in part by the National Science Foundation under Grant No. PHY-0854782. TM was supported in part by the Director, Office of Science, Office of Nuclear Physics, of the U.S. Department of Energy under grant numbers DE-FG02-05ER41368.  IZR  is supported by 
DOE DE-FG02-04ER41338 and FG02-06ER41449.  IZR also  acknowledges support from the Gordon and Betty Moore foundation, and thanks the the Caltech
theory group for its hospitality.  

\appendix

\section{SCET review}
\label{app1}
SCET is an effective field theory coupling soft and collinear degrees of freedom.  
Collinear degrees of freedom have light-cone momenta $(k^+,k^-,k_\perp)$ that scale as
\begin{equation}
p\sim Q(\lambda^2,1,\lambda), 
\end{equation}
where $Q$ is the large mass scale and $\lambda$ is the SCET expansion parameter. Soft modes have momenta that scale as
\begin{equation}
p\sim Q(\lambda,\lambda,\lambda), 
\end{equation}
while ultra-soft (usoft) modes scale as
\begin{equation}
p\sim Q(\lambda^2,\lambda^2,\lambda^2). 
\end{equation}
For example, if we are interested in describing the motion of a highly energetic particle with off-shellness $m^2$ in the light-like direction $n$, the collinear mode light-cone momentum will scale as $Q( \lambda^2,1,\lambda)$, where $\lambda \sim m/Q$ with $m \ll Q$ and the usoft mode momenta scale as $(\lambda^2,\lambda^2,\lambda^2)$. Which modes are present depends on the process.  In order to keep the parametrically different momenta separate we introduce a projection operator $\cP^\mu$ which projects out momentum of order $Q$ or $Q\lambda$. The derivative $\partial^\mu$ only operates on residual momenta $\sim Q \lambda^2$. 

SCET operators are constructed out of gauge invariant combinations of fields and collinear light-like Wilson lines~\cite{Bauer:2002nz} which for convenience are combined into a single field. For quarks that have large momentum component in the $n$ direction we define
\begin{equation} 
\chi_{n,\omega} \equiv \big[ \delta(\omega - \bnP) W^\dagger_n \xi_{n,\tilde p} \big] \,,
\end{equation}
where the quark field $\xi_{n,\tilde p}$ is labelled by the light-cone direction $n$ and by the large light-cone momentum components of $p$: $\tilde p =  \bn\!\cdot\!p n^\mu  /2 +p^\mu_\perp$. The operator $\bnP \equiv \bn \cdot {\cal P}$ acts on the quark field to project out the large light-cone momentum label: $\bnP \xi_{n,\tilde p} = \bn\!\cdot\!p  \, \xi_{n,\tilde p}$. The SCET collinear Wilson line is defined as
\begin{equation}
W^\dagger_n = \Bigg[ \sum_\textrm{\small perms} \textrm{exp} \bigg( -g \frac{1}{\bnP} \bn\cdot A_{n,q}\bigg)\Bigg]\,,
\end{equation}
where $A_{n,q}^\mu$ is the collinear gluon field.
The  gauge invariant field strength is
 \begin{equation}
 \big( {\cal G}_{n, \omega}\big)^{\mu, \nu} = -\frac{i}{g}\Big[ \delta(\omega - \bnP ) W^\dagger_n [i {\cal D}^\mu_n+g A^\mu_{n,q}, i {\cal D}^\nu_n+g A^\nu_{n,q}]W_n\Big]
\end{equation}
where 
\begin{equation}
i {\cal D}^\mu_n = \frac{n^\mu}{2} \bnP + {\cal P}^\mu_\perp + \frac{\bn^\mu}{2}  n\cdot(i \partial + gA_{us})\,,
\end{equation}
with $A_{us}$ being the ultra-soft gluon field. Since $ \big( {\cal G}_{n, \omega}\big)^{\mu, \nu} $ is not homogeneous in the SCET power counting we project out the leading contribution by introducing the gauge invariant field $\bnP B_{n,\omega}^{\alpha} \equiv i  n_\nu  \big( {\cal G}_{n, \omega}\big)^{\mu, \nu} g^{\hspace{1ex}\alpha}_{\perp \mu}$, where $g^{\mu \nu}_\perp \equiv g^{\mu\nu} -n^\mu \bn^\nu/2 - \bn^\mu n^\nu/2$ projects out the components of a four vector that are perpendicular to $n$ and $\bn$. Collinear operators are constructed out of combinations of the gauge invariant fields above. For example, an operator that creates a quark and an anti-quark moving in opposite light-like directions is $\bar \chi_{\bn, \bar\omega} \Gamma \chi_{n,\omega}$, where $\Gamma$ is a direct product of a Dirac matrix, and a color matrix depending on the current producing the quark--anti-quark pair.  The form of $\Gamma$ is constrained by the symmetries of SCET and for this combination of fields is restricted to $ \Gamma= \{ \mathbbm{1}, T^a\} \otimes \{ 1, \gamma_5,\gamma_\perp^\mu \}$, where $ \gamma^\mu_\perp =\gamma_\nu g^{\mu \nu}_\perp$. Since the production of two back-to-back light-like quarks with large energy is associated with a current that has large invariant mass the operator in our example must be matched  onto QCD. This gives the matching coefficient which can depend on the large light-cone momentum components $\omega$ and $\bar\omega$. As a result the short-distance coefficient and operator are convoluted :
\begin{equation}
O_\textrm{\tiny QCD} \to \int d \omega d \bar\omega \, C(\omega,\bar\omega)  \bar \chi_{\bn, \bar\omega} \Gamma \chi_{n,\omega}\,.
\end{equation}
Furthermore, usoft modes can be decoupled from collinear modes in the SCET Lagrangian through a field redefinition \cite{Bauer:2001yt}
\bea
\label{fieldredef}
\xi_{n,p} & \to & Y_n \xi_{n,p} \nn \\
A^\mu_{n,q} & \to & Y_n A^\mu_{n,q} Y^\dagger_n \nn \\
W_n & \to & Y_n W_n  Y^\dagger_n \nn \\
\chi_{n,\omega} & \to & Y_n \chi_{n,\omega}\,,
\eea
where the collinear fields on the right no longer couple to usoft fields, and $Y_n$ is a path-ordered exponential of the usoft gluon field
\begin{equation}
Y_n(x) = \textrm{P exp}\bigg( ig \int_{-\infty}^0 ds \, n\!\cdot\! A_{us}(s n+x)\bigg) \,.
\end{equation}

\bibliography{DPFFref}

\end{document}

%% file: rgb.tex
\definecolor{AliceBlue}{rgb}{0.94,0.97,1.00}
\definecolor{AntiqueWhite1}{rgb}{1.00,0.94,0.86}
\definecolor{AntiqueWhite2}{rgb}{0.93,0.87,0.80}
\definecolor{AntiqueWhite3}{rgb}{0.80,0.75,0.69}
\definecolor{AntiqueWhite4}{rgb}{0.55,0.51,0.47}
\definecolor{AntiqueWhite}{rgb}{0.98,0.92,0.84}
\definecolor{BlanchedAlmond}{rgb}{1.00,0.92,0.80}
\definecolor{BlueViolet}{rgb}{0.54,0.17,0.89}
\definecolor{CadetBlue1}{rgb}{0.60,0.96,1.00}
\definecolor{CadetBlue2}{rgb}{0.56,0.90,0.93}
\definecolor{CadetBlue3}{rgb}{0.48,0.77,0.80}
\definecolor{CadetBlue4}{rgb}{0.33,0.53,0.55}
\definecolor{CadetBlue}{rgb}{0.37,0.62,0.63}
\definecolor{CornflowerBlue}{rgb}{0.39,0.58,0.93}
\definecolor{DarkBlue}{rgb}{0.00,0.00,0.55}
\definecolor{DarkCyan}{rgb}{0.00,0.55,0.55}
\definecolor{DarkGoldenrod1}{rgb}{1.00,0.73,0.06}
\definecolor{DarkGoldenrod2}{rgb}{0.93,0.68,0.05}
\definecolor{DarkGoldenrod3}{rgb}{0.80,0.58,0.05}
\definecolor{DarkGoldenrod4}{rgb}{0.55,0.40,0.03}
\definecolor{DarkGoldenrod}{rgb}{0.72,0.53,0.04}
\definecolor{DarkGray}{rgb}{0.66,0.66,0.66}
\definecolor{DarkGreen}{rgb}{0.00,0.39,0.00}
\definecolor{DarkGrey}{rgb}{0.66,0.66,0.66}
\definecolor{DarkKhaki}{rgb}{0.74,0.72,0.42}
\definecolor{DarkMagenta}{rgb}{0.55,0.00,0.55}
\definecolor{DarkOliveGreen1}{rgb}{0.79,1.00,0.44}
\definecolor{DarkOliveGreen2}{rgb}{0.74,0.93,0.41}
\definecolor{DarkOliveGreen3}{rgb}{0.64,0.80,0.35}
\definecolor{DarkOliveGreen4}{rgb}{0.43,0.55,0.24}
\definecolor{DarkOliveGreen}{rgb}{0.33,0.42,0.18}
\definecolor{DarkOrange1}{rgb}{1.00,0.50,0.00}
\definecolor{DarkOrange2}{rgb}{0.93,0.46,0.00}
\definecolor{DarkOrange3}{rgb}{0.80,0.40,0.00}
\definecolor{DarkOrange4}{rgb}{0.55,0.27,0.00}
\definecolor{DarkOrange}{rgb}{1.00,0.55,0.00}
\definecolor{DarkOrchid1}{rgb}{0.75,0.24,1.00}
\definecolor{DarkOrchid2}{rgb}{0.70,0.23,0.93}
\definecolor{DarkOrchid3}{rgb}{0.60,0.20,0.80}
\definecolor{DarkOrchid4}{rgb}{0.41,0.13,0.55}
\definecolor{DarkOrchid}{rgb}{0.60,0.20,0.80}
\definecolor{DarkRed}{rgb}{0.55,0.00,0.00}
\definecolor{DarkSalmon}{rgb}{0.91,0.59,0.48}
\definecolor{DarkSeaGreen1}{rgb}{0.76,1.00,0.76}
\definecolor{DarkSeaGreen2}{rgb}{0.71,0.93,0.71}
\definecolor{DarkSeaGreen3}{rgb}{0.61,0.80,0.61}
\definecolor{DarkSeaGreen4}{rgb}{0.41,0.55,0.41}
\definecolor{DarkSeaGreen}{rgb}{0.56,0.74,0.56}
\definecolor{DarkSlateBlue}{rgb}{0.28,0.24,0.55}
\definecolor{DarkSlateGray1}{rgb}{0.59,1.00,1.00}
\definecolor{DarkSlateGray2}{rgb}{0.55,0.93,0.93}
\definecolor{DarkSlateGray3}{rgb}{0.47,0.80,0.80}
\definecolor{DarkSlateGray4}{rgb}{0.32,0.55,0.55}
\definecolor{DarkSlateGray}{rgb}{0.18,0.31,0.31}
\definecolor{DarkSlateGrey}{rgb}{0.18,0.31,0.31}
\definecolor{DarkTurquoise}{rgb}{0.00,0.81,0.82}
\definecolor{DarkViolet}{rgb}{0.58,0.00,0.83}
\definecolor{DeepPink1}{rgb}{1.00,0.08,0.58}
\definecolor{DeepPink2}{rgb}{0.93,0.07,0.54}
\definecolor{DeepPink3}{rgb}{0.80,0.06,0.46}
\definecolor{DeepPink4}{rgb}{0.55,0.04,0.31}
\definecolor{DeepPink}{rgb}{1.00,0.08,0.58}
\definecolor{DeepSkyBlue1}{rgb}{0.00,0.75,1.00}
\definecolor{DeepSkyBlue2}{rgb}{0.00,0.70,0.93}
\definecolor{DeepSkyBlue3}{rgb}{0.00,0.60,0.80}
\definecolor{DeepSkyBlue4}{rgb}{0.00,0.41,0.55}
\definecolor{DeepSkyBlue}{rgb}{0.00,0.75,1.00}
\definecolor{DimGray}{rgb}{0.41,0.41,0.41}
\definecolor{DimGrey}{rgb}{0.41,0.41,0.41}
\definecolor{DodgerBlue1}{rgb}{0.12,0.56,1.00}
\definecolor{DodgerBlue2}{rgb}{0.11,0.53,0.93}
\definecolor{DodgerBlue3}{rgb}{0.09,0.45,0.80}
\definecolor{DodgerBlue4}{rgb}{0.06,0.31,0.55}
\definecolor{DodgerBlue}{rgb}{0.12,0.56,1.00}
\definecolor{FloralWhite}{rgb}{1.00,0.98,0.94}
\definecolor{ForestGreen}{rgb}{0.13,0.55,0.13}
\definecolor{GhostWhite}{rgb}{0.97,0.97,1.00}
\definecolor{GreenYellow}{rgb}{0.68,1.00,0.18}
\definecolor{HotPink1}{rgb}{1.00,0.43,0.71}
\definecolor{HotPink2}{rgb}{0.93,0.42,0.65}
\definecolor{HotPink3}{rgb}{0.80,0.38,0.56}
\definecolor{HotPink4}{rgb}{0.55,0.23,0.38}
\definecolor{HotPink}{rgb}{1.00,0.41,0.71}
\definecolor{IndianRed1}{rgb}{1.00,0.42,0.42}
\definecolor{IndianRed2}{rgb}{0.93,0.39,0.39}
\definecolor{IndianRed3}{rgb}{0.80,0.33,0.33}
\definecolor{IndianRed4}{rgb}{0.55,0.23,0.23}
\definecolor{IndianRed}{rgb}{0.80,0.36,0.36}
\definecolor{LavenderBlush1}{rgb}{1.00,0.94,0.96}
\definecolor{LavenderBlush2}{rgb}{0.93,0.88,0.90}
\definecolor{LavenderBlush3}{rgb}{0.80,0.76,0.77}
\definecolor{LavenderBlush4}{rgb}{0.55,0.51,0.53}
\definecolor{LavenderBlush}{rgb}{1.00,0.94,0.96}
\definecolor{LawnGreen}{rgb}{0.49,0.99,0.00}
\definecolor{LemonChiffon1}{rgb}{1.00,0.98,0.80}
\definecolor{LemonChiffon2}{rgb}{0.93,0.91,0.75}
\definecolor{LemonChiffon3}{rgb}{0.80,0.79,0.65}
\definecolor{LemonChiffon4}{rgb}{0.55,0.54,0.44}
\definecolor{LemonChiffon}{rgb}{1.00,0.98,0.80}
\definecolor{LightBlue1}{rgb}{0.75,0.94,1.00}
\definecolor{LightBlue2}{rgb}{0.70,0.87,0.93}
\definecolor{LightBlue3}{rgb}{0.60,0.75,0.80}
\definecolor{LightBlue4}{rgb}{0.41,0.51,0.55}
\definecolor{LightBlue}{rgb}{0.68,0.85,0.90}
\definecolor{LightCoral}{rgb}{0.94,0.50,0.50}
\definecolor{LightCyan1}{rgb}{0.88,1.00,1.00}
\definecolor{LightCyan2}{rgb}{0.82,0.93,0.93}
\definecolor{LightCyan3}{rgb}{0.71,0.80,0.80}
\definecolor{LightCyan4}{rgb}{0.48,0.55,0.55}
\definecolor{LightCyan}{rgb}{0.88,1.00,1.00}
\definecolor{LightGoldenrod1}{rgb}{1.00,0.93,0.55}
\definecolor{LightGoldenrod2}{rgb}{0.93,0.86,0.51}
\definecolor{LightGoldenrod3}{rgb}{0.80,0.75,0.44}
\definecolor{LightGoldenrod4}{rgb}{0.55,0.51,0.30}
\definecolor{LightGoldenrodYellow}{rgb}{0.98,0.98,0.82}
\definecolor{LightGoldenrod}{rgb}{0.93,0.87,0.51}
\definecolor{LightGray}{rgb}{0.83,0.83,0.83}
\definecolor{LightGreen}{rgb}{0.56,0.93,0.56}
\definecolor{LightGrey}{rgb}{0.83,0.83,0.83}
\definecolor{LightPink1}{rgb}{1.00,0.68,0.73}
\definecolor{LightPink2}{rgb}{0.93,0.64,0.68}
\definecolor{LightPink3}{rgb}{0.80,0.55,0.58}
\definecolor{LightPink4}{rgb}{0.55,0.37,0.40}
\definecolor{LightPink}{rgb}{1.00,0.71,0.76}
\definecolor{LightSalmon1}{rgb}{1.00,0.63,0.48}
\definecolor{LightSalmon2}{rgb}{0.93,0.58,0.45}
\definecolor{LightSalmon3}{rgb}{0.80,0.51,0.38}
\definecolor{LightSalmon4}{rgb}{0.55,0.34,0.26}
\definecolor{LightSalmon}{rgb}{1.00,0.63,0.48}
\definecolor{LightSeaGreen}{rgb}{0.13,0.70,0.67}
\definecolor{LightSkyBlue1}{rgb}{0.69,0.89,1.00}
\definecolor{LightSkyBlue2}{rgb}{0.64,0.83,0.93}
\definecolor{LightSkyBlue3}{rgb}{0.55,0.71,0.80}
\definecolor{LightSkyBlue4}{rgb}{0.38,0.48,0.55}
\definecolor{LightSkyBlue}{rgb}{0.53,0.81,0.98}
\definecolor{LightSlateBlue}{rgb}{0.52,0.44,1.00}
\definecolor{LightSlateGray}{rgb}{0.47,0.53,0.60}
\definecolor{LightSlateGrey}{rgb}{0.47,0.53,0.60}
\definecolor{LightSteelBlue1}{rgb}{0.79,0.88,1.00}
\definecolor{LightSteelBlue2}{rgb}{0.74,0.82,0.93}
\definecolor{LightSteelBlue3}{rgb}{0.64,0.71,0.80}
\definecolor{LightSteelBlue4}{rgb}{0.43,0.48,0.55}
\definecolor{LightSteelBlue}{rgb}{0.69,0.77,0.87}
\definecolor{LightYellow1}{rgb}{1.00,1.00,0.88}
\definecolor{LightYellow2}{rgb}{0.93,0.93,0.82}
\definecolor{LightYellow3}{rgb}{0.80,0.80,0.71}
\definecolor{LightYellow4}{rgb}{0.55,0.55,0.48}
\definecolor{LightYellow}{rgb}{1.00,1.00,0.88}
\definecolor{LimeGreen}{rgb}{0.20,0.80,0.20}
\definecolor{MediumAquamarine}{rgb}{0.40,0.80,0.67}
\definecolor{MediumBlue}{rgb}{0.00,0.00,0.80}
\definecolor{MediumOrchid1}{rgb}{0.88,0.40,1.00}
\definecolor{MediumOrchid2}{rgb}{0.82,0.37,0.93}
\definecolor{MediumOrchid3}{rgb}{0.71,0.32,0.80}
\definecolor{MediumOrchid4}{rgb}{0.48,0.22,0.55}
\definecolor{MediumOrchid}{rgb}{0.73,0.33,0.83}
\definecolor{MediumPurple1}{rgb}{0.67,0.51,1.00}
\definecolor{MediumPurple2}{rgb}{0.62,0.47,0.93}
\definecolor{MediumPurple3}{rgb}{0.54,0.41,0.80}
\definecolor{MediumPurple4}{rgb}{0.36,0.28,0.55}
\definecolor{MediumPurple}{rgb}{0.58,0.44,0.86}
\definecolor{MediumSeaGreen}{rgb}{0.24,0.70,0.44}
\definecolor{MediumSlateBlue}{rgb}{0.48,0.41,0.93}
\definecolor{MediumSpringGreen}{rgb}{0.00,0.98,0.60}
\definecolor{MediumTurquoise}{rgb}{0.28,0.82,0.80}
\definecolor{MediumVioletRed}{rgb}{0.78,0.08,0.52}
\definecolor{MidnightBlue}{rgb}{0.10,0.10,0.44}
\definecolor{MintCream}{rgb}{0.96,1.00,0.98}
\definecolor{MistyRose1}{rgb}{1.00,0.89,0.88}
\definecolor{MistyRose2}{rgb}{0.93,0.84,0.82}
\definecolor{MistyRose3}{rgb}{0.80,0.72,0.71}
\definecolor{MistyRose4}{rgb}{0.55,0.49,0.48}
\definecolor{MistyRose}{rgb}{1.00,0.89,0.88}
\definecolor{NavajoWhite1}{rgb}{1.00,0.87,0.68}
\definecolor{NavajoWhite2}{rgb}{0.93,0.81,0.63}
\definecolor{NavajoWhite3}{rgb}{0.80,0.70,0.55}
\definecolor{NavajoWhite4}{rgb}{0.55,0.47,0.37}
\definecolor{NavajoWhite}{rgb}{1.00,0.87,0.68}
\definecolor{NavyBlue}{rgb}{0.00,0.00,0.50}
\definecolor{OldLace}{rgb}{0.99,0.96,0.90}
\definecolor{OliveDrab1}{rgb}{0.75,1.00,0.24}
\definecolor{OliveDrab2}{rgb}{0.70,0.93,0.23}
\definecolor{OliveDrab3}{rgb}{0.60,0.80,0.20}
\definecolor{OliveDrab4}{rgb}{0.41,0.55,0.13}
\definecolor{OliveDrab}{rgb}{0.42,0.56,0.14}
\definecolor{OrangeRed1}{rgb}{1.00,0.27,0.00}
\definecolor{OrangeRed2}{rgb}{0.93,0.25,0.00}
\definecolor{OrangeRed3}{rgb}{0.80,0.22,0.00}
\definecolor{OrangeRed4}{rgb}{0.55,0.15,0.00}
\definecolor{OrangeRed}{rgb}{1.00,0.27,0.00}
\definecolor{PaleGoldenrod}{rgb}{0.93,0.91,0.67}
\definecolor{PaleGreen1}{rgb}{0.60,1.00,0.60}
\definecolor{PaleGreen2}{rgb}{0.56,0.93,0.56}
\definecolor{PaleGreen3}{rgb}{0.49,0.80,0.49}
\definecolor{PaleGreen4}{rgb}{0.33,0.55,0.33}
\definecolor{PaleGreen}{rgb}{0.60,0.98,0.60}
\definecolor{PaleTurquoise1}{rgb}{0.73,1.00,1.00}
\definecolor{PaleTurquoise2}{rgb}{0.68,0.93,0.93}
\definecolor{PaleTurquoise3}{rgb}{0.59,0.80,0.80}
\definecolor{PaleTurquoise4}{rgb}{0.40,0.55,0.55}
\definecolor{PaleTurquoise}{rgb}{0.69,0.93,0.93}
\definecolor{PaleVioletRed1}{rgb}{1.00,0.51,0.67}
\definecolor{PaleVioletRed2}{rgb}{0.93,0.47,0.62}
\definecolor{PaleVioletRed3}{rgb}{0.80,0.41,0.54}
\definecolor{PaleVioletRed4}{rgb}{0.55,0.28,0.36}
\definecolor{PaleVioletRed}{rgb}{0.86,0.44,0.58}
\definecolor{PapayaWhip}{rgb}{1.00,0.94,0.84}
\definecolor{PeachPuff1}{rgb}{1.00,0.85,0.73}
\definecolor{PeachPuff2}{rgb}{0.93,0.80,0.68}
\definecolor{PeachPuff3}{rgb}{0.80,0.69,0.58}
\definecolor{PeachPuff4}{rgb}{0.55,0.47,0.40}
\definecolor{PeachPuff}{rgb}{1.00,0.85,0.73}
\definecolor{PowderBlue}{rgb}{0.69,0.88,0.90}
\definecolor{RosyBrown1}{rgb}{1.00,0.76,0.76}
\definecolor{RosyBrown2}{rgb}{0.93,0.71,0.71}
\definecolor{RosyBrown3}{rgb}{0.80,0.61,0.61}
\definecolor{RosyBrown4}{rgb}{0.55,0.41,0.41}
\definecolor{RosyBrown}{rgb}{0.74,0.56,0.56}
\definecolor{RoyalBlue1}{rgb}{0.28,0.46,1.00}
\definecolor{RoyalBlue2}{rgb}{0.26,0.43,0.93}
\definecolor{RoyalBlue3}{rgb}{0.23,0.37,0.80}
\definecolor{RoyalBlue4}{rgb}{0.15,0.25,0.55}
\definecolor{RoyalBlue}{rgb}{0.25,0.41,0.88}
\definecolor{SaddleBrown}{rgb}{0.55,0.27,0.07}
\definecolor{SandyBrown}{rgb}{0.96,0.64,0.38}
\definecolor{SeaGreen1}{rgb}{0.33,1.00,0.62}
\definecolor{SeaGreen2}{rgb}{0.31,0.93,0.58}
\definecolor{SeaGreen3}{rgb}{0.26,0.80,0.50}
\definecolor{SeaGreen4}{rgb}{0.18,0.55,0.34}
\definecolor{SeaGreen}{rgb}{0.18,0.55,0.34}
\definecolor{SkyBlue1}{rgb}{0.53,0.81,1.00}
\definecolor{SkyBlue2}{rgb}{0.49,0.75,0.93}
\definecolor{SkyBlue3}{rgb}{0.42,0.65,0.80}
\definecolor{SkyBlue4}{rgb}{0.29,0.44,0.55}
\definecolor{SkyBlue}{rgb}{0.53,0.81,0.92}
\definecolor{SlateBlue1}{rgb}{0.51,0.44,1.00}
\definecolor{SlateBlue2}{rgb}{0.48,0.40,0.93}
\definecolor{SlateBlue3}{rgb}{0.41,0.35,0.80}
\definecolor{SlateBlue4}{rgb}{0.28,0.24,0.55}
\definecolor{SlateBlue}{rgb}{0.42,0.35,0.80}
\definecolor{SlateGray1}{rgb}{0.78,0.89,1.00}
\definecolor{SlateGray2}{rgb}{0.73,0.83,0.93}
\definecolor{SlateGray3}{rgb}{0.62,0.71,0.80}
\definecolor{SlateGray4}{rgb}{0.42,0.48,0.55}
\definecolor{SlateGray}{rgb}{0.44,0.50,0.56}
\definecolor{SlateGrey}{rgb}{0.44,0.50,0.56}
\definecolor{SpringGreen1}{rgb}{0.00,1.00,0.50}
\definecolor{SpringGreen2}{rgb}{0.00,0.93,0.46}
\definecolor{SpringGreen3}{rgb}{0.00,0.80,0.40}
\definecolor{SpringGreen4}{rgb}{0.00,0.55,0.27}
\definecolor{SpringGreen}{rgb}{0.00,1.00,0.50}
\definecolor{SteelBlue1}{rgb}{0.39,0.72,1.00}
\definecolor{SteelBlue2}{rgb}{0.36,0.67,0.93}
\definecolor{SteelBlue3}{rgb}{0.31,0.58,0.80}
\definecolor{SteelBlue4}{rgb}{0.21,0.39,0.55}
\definecolor{SteelBlue}{rgb}{0.27,0.51,0.71}
\definecolor{VioletRed1}{rgb}{1.00,0.24,0.59}
\definecolor{VioletRed2}{rgb}{0.93,0.23,0.55}
\definecolor{VioletRed3}{rgb}{0.80,0.20,0.47}
\definecolor{VioletRed4}{rgb}{0.55,0.13,0.32}
\definecolor{VioletRed}{rgb}{0.82,0.13,0.56}
\definecolor{WhiteSmoke}{rgb}{0.96,0.96,0.96}
\definecolor{YellowGreen}{rgb}{0.60,0.80,0.20}
\definecolor{aliceblue}{rgb}{0.94,0.97,1.00}
\definecolor{antiquewhite}{rgb}{0.98,0.92,0.84}
\definecolor{aquamarine1}{rgb}{0.50,1.00,0.83}
\definecolor{aquamarine2}{rgb}{0.46,0.93,0.78}
\definecolor{aquamarine3}{rgb}{0.40,0.80,0.67}
\definecolor{aquamarine4}{rgb}{0.27,0.55,0.45}
\definecolor{aquamarine}{rgb}{0.50,1.00,0.83}
\definecolor{azure1}{rgb}{0.94,1.00,1.00}
\definecolor{azure2}{rgb}{0.88,0.93,0.93}
\definecolor{azure3}{rgb}{0.76,0.80,0.80}
\definecolor{azure4}{rgb}{0.51,0.55,0.55}
\definecolor{azure}{rgb}{0.94,1.00,1.00}
\definecolor{beige}{rgb}{0.96,0.96,0.86}
\definecolor{bisque1}{rgb}{1.00,0.89,0.77}
\definecolor{bisque2}{rgb}{0.93,0.84,0.72}
\definecolor{bisque3}{rgb}{0.80,0.72,0.62}
\definecolor{bisque4}{rgb}{0.55,0.49,0.42}
\definecolor{bisque}{rgb}{1.00,0.89,0.77}
\definecolor{black}{rgb}{0.00,0.00,0.00}
\definecolor{blanchedalmond}{rgb}{1.00,0.92,0.80}
\definecolor{blue1}{rgb}{0.00,0.00,1.00}
\definecolor{blue2}{rgb}{0.00,0.00,0.93}
\definecolor{blue3}{rgb}{0.00,0.00,0.80}
\definecolor{blue4}{rgb}{0.00,0.00,0.55}
\definecolor{blueviolet}{rgb}{0.54,0.17,0.89}
\definecolor{blue}{rgb}{0.00,0.00,1.00}
\definecolor{brown1}{rgb}{1.00,0.25,0.25}
\definecolor{brown2}{rgb}{0.93,0.23,0.23}
\definecolor{brown3}{rgb}{0.80,0.20,0.20}
\definecolor{brown4}{rgb}{0.55,0.14,0.14}
\definecolor{brown}{rgb}{0.65,0.16,0.16}
\definecolor{burlywood1}{rgb}{1.00,0.83,0.61}
\definecolor{burlywood2}{rgb}{0.93,0.77,0.57}
\definecolor{burlywood3}{rgb}{0.80,0.67,0.49}
\definecolor{burlywood4}{rgb}{0.55,0.45,0.33}
\definecolor{burlywood}{rgb}{0.87,0.72,0.53}
\definecolor{cadetblue}{rgb}{0.37,0.62,0.63}
\definecolor{chartreuse1}{rgb}{0.50,1.00,0.00}
\definecolor{chartreuse2}{rgb}{0.46,0.93,0.00}
\definecolor{chartreuse3}{rgb}{0.40,0.80,0.00}
\definecolor{chartreuse4}{rgb}{0.27,0.55,0.00}
\definecolor{chartreuse}{rgb}{0.50,1.00,0.00}
\definecolor{chocolate1}{rgb}{1.00,0.50,0.14}
\definecolor{chocolate2}{rgb}{0.93,0.46,0.13}
\definecolor{chocolate3}{rgb}{0.80,0.40,0.11}
\definecolor{chocolate4}{rgb}{0.55,0.27,0.07}
\definecolor{chocolate}{rgb}{0.82,0.41,0.12}
\definecolor{coral1}{rgb}{1.00,0.45,0.34}
\definecolor{coral2}{rgb}{0.93,0.42,0.31}
\definecolor{coral3}{rgb}{0.80,0.36,0.27}
\definecolor{coral4}{rgb}{0.55,0.24,0.18}
\definecolor{coral}{rgb}{1.00,0.50,0.31}
\definecolor{cornflowerblue}{rgb}{0.39,0.58,0.93}
\definecolor{cornsilk1}{rgb}{1.00,0.97,0.86}
\definecolor{cornsilk2}{rgb}{0.93,0.91,0.80}
\definecolor{cornsilk3}{rgb}{0.80,0.78,0.69}
\definecolor{cornsilk4}{rgb}{0.55,0.53,0.47}
\definecolor{cornsilk}{rgb}{1.00,0.97,0.86}
\definecolor{cyan1}{rgb}{0.00,1.00,1.00}
\definecolor{cyan2}{rgb}{0.00,0.93,0.93}
\definecolor{cyan3}{rgb}{0.00,0.80,0.80}
\definecolor{cyan4}{rgb}{0.00,0.55,0.55}
\definecolor{cyan}{rgb}{0.00,1.00,1.00}
\definecolor{darkblue}{rgb}{0.00,0.00,0.55}
\definecolor{darkcyan}{rgb}{0.00,0.55,0.55}
\definecolor{darkgoldenrod}{rgb}{0.72,0.53,0.04}
\definecolor{darkgray}{rgb}{0.66,0.66,0.66}
\definecolor{darkgreen}{rgb}{0.00,0.39,0.00}
\definecolor{darkgrey}{rgb}{0.66,0.66,0.66}
\definecolor{darkkhaki}{rgb}{0.74,0.72,0.42}
\definecolor{darkmagenta}{rgb}{0.55,0.00,0.55}
\definecolor{darkolive}{rgb}{0.33,0.42,0.18}
\definecolor{darkorange}{rgb}{1.00,0.55,0.00}
\definecolor{darkorchid}{rgb}{0.60,0.20,0.80}
\definecolor{darkred}{rgb}{0.55,0.00,0.00}
\definecolor{darksalmon}{rgb}{0.91,0.59,0.48}
\definecolor{darksea}{rgb}{0.56,0.74,0.56}
\definecolor{darkslate}{rgb}{0.18,0.31,0.31}
\definecolor{darkslate}{rgb}{0.18,0.31,0.31}
\definecolor{darkslate}{rgb}{0.28,0.24,0.55}
\definecolor{darkturquoise}{rgb}{0.00,0.81,0.82}
\definecolor{darkviolet}{rgb}{0.58,0.00,0.83}
\definecolor{deeppink}{rgb}{1.00,0.08,0.58}
\definecolor{deepsky}{rgb}{0.00,0.75,1.00}
\definecolor{dimgray}{rgb}{0.41,0.41,0.41}
\definecolor{dimgrey}{rgb}{0.41,0.41,0.41}
\definecolor{dodgerblue}{rgb}{0.12,0.56,1.00}
\definecolor{firebrick1}{rgb}{1.00,0.19,0.19}
\definecolor{firebrick2}{rgb}{0.93,0.17,0.17}
\definecolor{firebrick3}{rgb}{0.80,0.15,0.15}
\definecolor{firebrick4}{rgb}{0.55,0.10,0.10}
\definecolor{firebrick}{rgb}{0.70,0.13,0.13}
\definecolor{floralwhite}{rgb}{1.00,0.98,0.94}
\definecolor{forestgreen}{rgb}{0.13,0.55,0.13}
\definecolor{gainsboro}{rgb}{0.86,0.86,0.86}
\definecolor{ghostwhite}{rgb}{0.97,0.97,1.00}
\definecolor{gold1}{rgb}{1.00,0.84,0.00}
\definecolor{gold2}{rgb}{0.93,0.79,0.00}
\definecolor{gold3}{rgb}{0.80,0.68,0.00}
\definecolor{gold4}{rgb}{0.55,0.46,0.00}
\definecolor{goldenrod1}{rgb}{1.00,0.76,0.15}
\definecolor{goldenrod2}{rgb}{0.93,0.71,0.13}
\definecolor{goldenrod3}{rgb}{0.80,0.61,0.11}
\definecolor{goldenrod4}{rgb}{0.55,0.41,0.08}
\definecolor{goldenrod}{rgb}{0.85,0.65,0.13}
\definecolor{gold}{rgb}{1.00,0.84,0.00}
\definecolor{gray0}{rgb}{0.00,0.00,0.00}
\definecolor{gray100}{rgb}{1.00,1.00,1.00}
\definecolor{gray10}{rgb}{0.10,0.10,0.10}
\definecolor{gray11}{rgb}{0.11,0.11,0.11}
\definecolor{gray12}{rgb}{0.12,0.12,0.12}
\definecolor{gray13}{rgb}{0.13,0.13,0.13}
\definecolor{gray14}{rgb}{0.14,0.14,0.14}
\definecolor{gray15}{rgb}{0.15,0.15,0.15}
\definecolor{gray16}{rgb}{0.16,0.16,0.16}
\definecolor{gray17}{rgb}{0.17,0.17,0.17}
\definecolor{gray18}{rgb}{0.18,0.18,0.18}
\definecolor{gray19}{rgb}{0.19,0.19,0.19}
\definecolor{gray1}{rgb}{0.01,0.01,0.01}
\definecolor{gray20}{rgb}{0.20,0.20,0.20}
\definecolor{gray21}{rgb}{0.21,0.21,0.21}
\definecolor{gray22}{rgb}{0.22,0.22,0.22}
\definecolor{gray23}{rgb}{0.23,0.23,0.23}
\definecolor{gray24}{rgb}{0.24,0.24,0.24}
\definecolor{gray25}{rgb}{0.25,0.25,0.25}
\definecolor{gray26}{rgb}{0.26,0.26,0.26}
\definecolor{gray27}{rgb}{0.27,0.27,0.27}
\definecolor{gray28}{rgb}{0.28,0.28,0.28}
\definecolor{gray29}{rgb}{0.29,0.29,0.29}
\definecolor{gray2}{rgb}{0.02,0.02,0.02}
\definecolor{gray30}{rgb}{0.30,0.30,0.30}
\definecolor{gray31}{rgb}{0.31,0.31,0.31}
\definecolor{gray32}{rgb}{0.32,0.32,0.32}
\definecolor{gray33}{rgb}{0.33,0.33,0.33}
\definecolor{gray34}{rgb}{0.34,0.34,0.34}
\definecolor{gray35}{rgb}{0.35,0.35,0.35}
\definecolor{gray36}{rgb}{0.36,0.36,0.36}
\definecolor{gray37}{rgb}{0.37,0.37,0.37}
\definecolor{gray38}{rgb}{0.38,0.38,0.38}
\definecolor{gray39}{rgb}{0.39,0.39,0.39}
\definecolor{gray3}{rgb}{0.03,0.03,0.03}
\definecolor{gray40}{rgb}{0.40,0.40,0.40}
\definecolor{gray41}{rgb}{0.41,0.41,0.41}
\definecolor{gray42}{rgb}{0.42,0.42,0.42}
\definecolor{gray43}{rgb}{0.43,0.43,0.43}
\definecolor{gray44}{rgb}{0.44,0.44,0.44}
\definecolor{gray45}{rgb}{0.45,0.45,0.45}
\definecolor{gray46}{rgb}{0.46,0.46,0.46}
\definecolor{gray47}{rgb}{0.47,0.47,0.47}
\definecolor{gray48}{rgb}{0.48,0.48,0.48}
\definecolor{gray49}{rgb}{0.49,0.49,0.49}
\definecolor{gray4}{rgb}{0.04,0.04,0.04}
\definecolor{gray50}{rgb}{0.50,0.50,0.50}
\definecolor{gray51}{rgb}{0.51,0.51,0.51}
\definecolor{gray52}{rgb}{0.52,0.52,0.52}
\definecolor{gray53}{rgb}{0.53,0.53,0.53}
\definecolor{gray54}{rgb}{0.54,0.54,0.54}
\definecolor{gray55}{rgb}{0.55,0.55,0.55}
\definecolor{gray56}{rgb}{0.56,0.56,0.56}
\definecolor{gray57}{rgb}{0.57,0.57,0.57}
\definecolor{gray58}{rgb}{0.58,0.58,0.58}
\definecolor{gray59}{rgb}{0.59,0.59,0.59}
\definecolor{gray5}{rgb}{0.05,0.05,0.05}
\definecolor{gray60}{rgb}{0.60,0.60,0.60}
\definecolor{gray61}{rgb}{0.61,0.61,0.61}
\definecolor{gray62}{rgb}{0.62,0.62,0.62}
\definecolor{gray63}{rgb}{0.63,0.63,0.63}
\definecolor{gray64}{rgb}{0.64,0.64,0.64}
\definecolor{gray65}{rgb}{0.65,0.65,0.65}
\definecolor{gray66}{rgb}{0.66,0.66,0.66}
\definecolor{gray67}{rgb}{0.67,0.67,0.67}
\definecolor{gray68}{rgb}{0.68,0.68,0.68}
\definecolor{gray69}{rgb}{0.69,0.69,0.69}
\definecolor{gray6}{rgb}{0.06,0.06,0.06}
\definecolor{gray70}{rgb}{0.70,0.70,0.70}
\definecolor{gray71}{rgb}{0.71,0.71,0.71}
\definecolor{gray72}{rgb}{0.72,0.72,0.72}
\definecolor{gray73}{rgb}{0.73,0.73,0.73}
\definecolor{gray74}{rgb}{0.74,0.74,0.74}
\definecolor{gray75}{rgb}{0.75,0.75,0.75}
\definecolor{gray76}{rgb}{0.76,0.76,0.76}
\definecolor{gray77}{rgb}{0.77,0.77,0.77}
\definecolor{gray78}{rgb}{0.78,0.78,0.78}
\definecolor{gray79}{rgb}{0.79,0.79,0.79}
\definecolor{gray7}{rgb}{0.07,0.07,0.07}
\definecolor{gray80}{rgb}{0.80,0.80,0.80}
\definecolor{gray81}{rgb}{0.81,0.81,0.81}
\definecolor{gray82}{rgb}{0.82,0.82,0.82}
\definecolor{gray83}{rgb}{0.83,0.83,0.83}
\definecolor{gray84}{rgb}{0.84,0.84,0.84}
\definecolor{gray85}{rgb}{0.85,0.85,0.85}
\definecolor{gray86}{rgb}{0.86,0.86,0.86}
\definecolor{gray87}{rgb}{0.87,0.87,0.87}
\definecolor{gray88}{rgb}{0.88,0.88,0.88}
\definecolor{gray89}{rgb}{0.89,0.89,0.89}
\definecolor{gray8}{rgb}{0.08,0.08,0.08}
\definecolor{gray90}{rgb}{0.90,0.90,0.90}
\definecolor{gray91}{rgb}{0.91,0.91,0.91}
\definecolor{gray92}{rgb}{0.92,0.92,0.92}
\definecolor{gray93}{rgb}{0.93,0.93,0.93}
\definecolor{gray94}{rgb}{0.94,0.94,0.94}
\definecolor{gray95}{rgb}{0.95,0.95,0.95}
\definecolor{gray96}{rgb}{0.96,0.96,0.96}
\definecolor{gray97}{rgb}{0.97,0.97,0.97}
\definecolor{gray98}{rgb}{0.98,0.98,0.98}
\definecolor{gray99}{rgb}{0.99,0.99,0.99}
\definecolor{gray9}{rgb}{0.09,0.09,0.09}
\definecolor{gray}{rgb}{0.75,0.75,0.75}
\definecolor{green1}{rgb}{0.00,1.00,0.00}
\definecolor{green2}{rgb}{0.00,0.93,0.00}
\definecolor{green3}{rgb}{0.00,0.80,0.00}
\definecolor{green4}{rgb}{0.00,0.55,0.00}
\definecolor{greenyellow}{rgb}{0.68,1.00,0.18}
\definecolor{green}{rgb}{0.00,1.00,0.00}
\definecolor{grey0}{rgb}{0.00,0.00,0.00}
\definecolor{grey100}{rgb}{1.00,1.00,1.00}
\definecolor{grey10}{rgb}{0.10,0.10,0.10}
\definecolor{grey11}{rgb}{0.11,0.11,0.11}
\definecolor{grey12}{rgb}{0.12,0.12,0.12}
\definecolor{grey13}{rgb}{0.13,0.13,0.13}
\definecolor{grey14}{rgb}{0.14,0.14,0.14}
\definecolor{grey15}{rgb}{0.15,0.15,0.15}
\definecolor{grey16}{rgb}{0.16,0.16,0.16}
\definecolor{grey17}{rgb}{0.17,0.17,0.17}
\definecolor{grey18}{rgb}{0.18,0.18,0.18}
\definecolor{grey19}{rgb}{0.19,0.19,0.19}
\definecolor{grey1}{rgb}{0.01,0.01,0.01}
\definecolor{grey20}{rgb}{0.20,0.20,0.20}
\definecolor{grey21}{rgb}{0.21,0.21,0.21}
\definecolor{grey22}{rgb}{0.22,0.22,0.22}
\definecolor{grey23}{rgb}{0.23,0.23,0.23}
\definecolor{grey24}{rgb}{0.24,0.24,0.24}
\definecolor{grey25}{rgb}{0.25,0.25,0.25}
\definecolor{grey26}{rgb}{0.26,0.26,0.26}
\definecolor{grey27}{rgb}{0.27,0.27,0.27}
\definecolor{grey28}{rgb}{0.28,0.28,0.28}
\definecolor{grey29}{rgb}{0.29,0.29,0.29}
\definecolor{grey2}{rgb}{0.02,0.02,0.02}
\definecolor{grey30}{rgb}{0.30,0.30,0.30}
\definecolor{grey31}{rgb}{0.31,0.31,0.31}
\definecolor{grey32}{rgb}{0.32,0.32,0.32}
\definecolor{grey33}{rgb}{0.33,0.33,0.33}
\definecolor{grey34}{rgb}{0.34,0.34,0.34}
\definecolor{grey35}{rgb}{0.35,0.35,0.35}
\definecolor{grey36}{rgb}{0.36,0.36,0.36}
\definecolor{grey37}{rgb}{0.37,0.37,0.37}
\definecolor{grey38}{rgb}{0.38,0.38,0.38}
\definecolor{grey39}{rgb}{0.39,0.39,0.39}
\definecolor{grey3}{rgb}{0.03,0.03,0.03}
\definecolor{grey40}{rgb}{0.40,0.40,0.40}
\definecolor{grey41}{rgb}{0.41,0.41,0.41}
\definecolor{grey42}{rgb}{0.42,0.42,0.42}
\definecolor{grey43}{rgb}{0.43,0.43,0.43}
\definecolor{grey44}{rgb}{0.44,0.44,0.44}
\definecolor{grey45}{rgb}{0.45,0.45,0.45}
\definecolor{grey46}{rgb}{0.46,0.46,0.46}
\definecolor{grey47}{rgb}{0.47,0.47,0.47}
\definecolor{grey48}{rgb}{0.48,0.48,0.48}
\definecolor{grey49}{rgb}{0.49,0.49,0.49}
\definecolor{grey4}{rgb}{0.04,0.04,0.04}
\definecolor{grey50}{rgb}{0.50,0.50,0.50}
\definecolor{grey51}{rgb}{0.51,0.51,0.51}
\definecolor{grey52}{rgb}{0.52,0.52,0.52}
\definecolor{grey53}{rgb}{0.53,0.53,0.53}
\definecolor{grey54}{rgb}{0.54,0.54,0.54}
\definecolor{grey55}{rgb}{0.55,0.55,0.55}
\definecolor{grey56}{rgb}{0.56,0.56,0.56}
\definecolor{grey57}{rgb}{0.57,0.57,0.57}
\definecolor{grey58}{rgb}{0.58,0.58,0.58}
\definecolor{grey59}{rgb}{0.59,0.59,0.59}
\definecolor{grey5}{rgb}{0.05,0.05,0.05}
\definecolor{grey60}{rgb}{0.60,0.60,0.60}
\definecolor{grey61}{rgb}{0.61,0.61,0.61}
\definecolor{grey62}{rgb}{0.62,0.62,0.62}
\definecolor{grey63}{rgb}{0.63,0.63,0.63}
\definecolor{grey64}{rgb}{0.64,0.64,0.64}
\definecolor{grey65}{rgb}{0.65,0.65,0.65}
\definecolor{grey66}{rgb}{0.66,0.66,0.66}
\definecolor{grey67}{rgb}{0.67,0.67,0.67}
\definecolor{grey68}{rgb}{0.68,0.68,0.68}
\definecolor{grey69}{rgb}{0.69,0.69,0.69}
\definecolor{grey6}{rgb}{0.06,0.06,0.06}
\definecolor{grey70}{rgb}{0.70,0.70,0.70}
\definecolor{grey71}{rgb}{0.71,0.71,0.71}
\definecolor{grey72}{rgb}{0.72,0.72,0.72}
\definecolor{grey73}{rgb}{0.73,0.73,0.73}
\definecolor{grey74}{rgb}{0.74,0.74,0.74}
\definecolor{grey75}{rgb}{0.75,0.75,0.75}
\definecolor{grey76}{rgb}{0.76,0.76,0.76}
\definecolor{grey77}{rgb}{0.77,0.77,0.77}
\definecolor{grey78}{rgb}{0.78,0.78,0.78}
\definecolor{grey79}{rgb}{0.79,0.79,0.79}
\definecolor{grey7}{rgb}{0.07,0.07,0.07}
\definecolor{grey80}{rgb}{0.80,0.80,0.80}
\definecolor{grey81}{rgb}{0.81,0.81,0.81}
\definecolor{grey82}{rgb}{0.82,0.82,0.82}
\definecolor{grey83}{rgb}{0.83,0.83,0.83}
\definecolor{grey84}{rgb}{0.84,0.84,0.84}
\definecolor{grey85}{rgb}{0.85,0.85,0.85}
\definecolor{grey86}{rgb}{0.86,0.86,0.86}
\definecolor{grey87}{rgb}{0.87,0.87,0.87}
\definecolor{grey88}{rgb}{0.88,0.88,0.88}
\definecolor{grey89}{rgb}{0.89,0.89,0.89}
\definecolor{grey8}{rgb}{0.08,0.08,0.08}
\definecolor{grey90}{rgb}{0.90,0.90,0.90}
\definecolor{grey91}{rgb}{0.91,0.91,0.91}
\definecolor{grey92}{rgb}{0.92,0.92,0.92}
\definecolor{grey93}{rgb}{0.93,0.93,0.93}
\definecolor{grey94}{rgb}{0.94,0.94,0.94}
\definecolor{grey95}{rgb}{0.95,0.95,0.95}
\definecolor{grey96}{rgb}{0.96,0.96,0.96}
\definecolor{grey97}{rgb}{0.97,0.97,0.97}
\definecolor{grey98}{rgb}{0.98,0.98,0.98}
\definecolor{grey99}{rgb}{0.99,0.99,0.99}
\definecolor{grey9}{rgb}{0.09,0.09,0.09}
\definecolor{grey}{rgb}{0.75,0.75,0.75}
\definecolor{honeydew1}{rgb}{0.94,1.00,0.94}
\definecolor{honeydew2}{rgb}{0.88,0.93,0.88}
\definecolor{honeydew3}{rgb}{0.76,0.80,0.76}
\definecolor{honeydew4}{rgb}{0.51,0.55,0.51}
\definecolor{honeydew}{rgb}{0.94,1.00,0.94}
\definecolor{hotpink}{rgb}{1.00,0.41,0.71}
\definecolor{indianred}{rgb}{0.80,0.36,0.36}
\definecolor{ivory1}{rgb}{1.00,1.00,0.94}
\definecolor{ivory2}{rgb}{0.93,0.93,0.88}
\definecolor{ivory3}{rgb}{0.80,0.80,0.76}
\definecolor{ivory4}{rgb}{0.55,0.55,0.51}
\definecolor{ivory}{rgb}{1.00,1.00,0.94}
\definecolor{khaki1}{rgb}{1.00,0.96,0.56}
\definecolor{khaki2}{rgb}{0.93,0.90,0.52}
\definecolor{khaki3}{rgb}{0.80,0.78,0.45}
\definecolor{khaki4}{rgb}{0.55,0.53,0.31}
\definecolor{khaki}{rgb}{0.94,0.90,0.55}
\definecolor{lavenderblush}{rgb}{1.00,0.94,0.96}
\definecolor{lavender}{rgb}{0.90,0.90,0.98}
\definecolor{lawngreen}{rgb}{0.49,0.99,0.00}
\definecolor{lemonchiffon}{rgb}{1.00,0.98,0.80}
\definecolor{lightblue}{rgb}{0.68,0.85,0.90}
\definecolor{lightcoral}{rgb}{0.94,0.50,0.50}
\definecolor{lightcyan}{rgb}{0.88,1.00,1.00}
\definecolor{lightgoldenrod}{rgb}{0.93,0.87,0.51}
\definecolor{lightgoldenrod}{rgb}{0.98,0.98,0.82}
\definecolor{lightgray}{rgb}{0.83,0.83,0.83}
\definecolor{lightgreen}{rgb}{0.56,0.93,0.56}
\definecolor{lightgrey}{rgb}{0.83,0.83,0.83}
\definecolor{lightpink}{rgb}{1.00,0.71,0.76}
\definecolor{lightsalmon}{rgb}{1.00,0.63,0.48}
\definecolor{lightsea}{rgb}{0.13,0.70,0.67}
\definecolor{lightsky}{rgb}{0.53,0.81,0.98}
\definecolor{lightslate}{rgb}{0.47,0.53,0.60}
\definecolor{lightslate}{rgb}{0.47,0.53,0.60}
\definecolor{lightslate}{rgb}{0.52,0.44,1.00}
\definecolor{lightsteel}{rgb}{0.69,0.77,0.87}
\definecolor{lightyellow}{rgb}{1.00,1.00,0.88}
\definecolor{limegreen}{rgb}{0.20,0.80,0.20}
\definecolor{linen}{rgb}{0.98,0.94,0.90}
\definecolor{magenta1}{rgb}{1.00,0.00,1.00}
\definecolor{magenta2}{rgb}{0.93,0.00,0.93}
\definecolor{magenta3}{rgb}{0.80,0.00,0.80}
\definecolor{magenta4}{rgb}{0.55,0.00,0.55}
\definecolor{magenta}{rgb}{1.00,0.00,1.00}
\definecolor{maroon1}{rgb}{1.00,0.20,0.70}
\definecolor{maroon2}{rgb}{0.93,0.19,0.65}
\definecolor{maroon3}{rgb}{0.80,0.16,0.56}
\definecolor{maroon4}{rgb}{0.55,0.11,0.38}
\definecolor{maroon}{rgb}{0.69,0.19,0.38}
\definecolor{mediumaquamarine}{rgb}{0.40,0.80,0.67}
\definecolor{mediumblue}{rgb}{0.00,0.00,0.80}
\definecolor{mediumorchid}{rgb}{0.73,0.33,0.83}
\definecolor{mediumpurple}{rgb}{0.58,0.44,0.86}
\definecolor{mediumsea}{rgb}{0.24,0.70,0.44}
\definecolor{mediumslate}{rgb}{0.48,0.41,0.93}
\definecolor{mediumspring}{rgb}{0.00,0.98,0.60}
\definecolor{mediumturquoise}{rgb}{0.28,0.82,0.80}
\definecolor{mediumviolet}{rgb}{0.78,0.08,0.52}
\definecolor{midnightblue}{rgb}{0.10,0.10,0.44}
\definecolor{mintcream}{rgb}{0.96,1.00,0.98}
\definecolor{mistyrose}{rgb}{1.00,0.89,0.88}
\definecolor{moccasin}{rgb}{1.00,0.89,0.71}
\definecolor{navajowhite}{rgb}{1.00,0.87,0.68}
\definecolor{navyblue}{rgb}{0.00,0.00,0.50}
\definecolor{navy}{rgb}{0.00,0.00,0.50}
\definecolor{oldlace}{rgb}{0.99,0.96,0.90}
\definecolor{olivedrab}{rgb}{0.42,0.56,0.14}
\definecolor{orange1}{rgb}{1.00,0.65,0.00}
\definecolor{orange2}{rgb}{0.93,0.60,0.00}
\definecolor{orange3}{rgb}{0.80,0.52,0.00}
\definecolor{orange4}{rgb}{0.55,0.35,0.00}
\definecolor{orangered}{rgb}{1.00,0.27,0.00}
\definecolor{orange}{rgb}{1.00,0.65,0.00}
\definecolor{orchid1}{rgb}{1.00,0.51,0.98}
\definecolor{orchid2}{rgb}{0.93,0.48,0.91}
\definecolor{orchid3}{rgb}{0.80,0.41,0.79}
\definecolor{orchid4}{rgb}{0.55,0.28,0.54}
\definecolor{orchid}{rgb}{0.85,0.44,0.84}
\definecolor{palegoldenrod}{rgb}{0.93,0.91,0.67}
\definecolor{palegreen}{rgb}{0.60,0.98,0.60}
\definecolor{paleturquoise}{rgb}{0.69,0.93,0.93}
\definecolor{paleviolet}{rgb}{0.86,0.44,0.58}
\definecolor{papayawhip}{rgb}{1.00,0.94,0.84}
\definecolor{peachpuff}{rgb}{1.00,0.85,0.73}
\definecolor{peru}{rgb}{0.80,0.52,0.25}
\definecolor{pink1}{rgb}{1.00,0.71,0.77}
\definecolor{pink2}{rgb}{0.93,0.66,0.72}
\definecolor{pink3}{rgb}{0.80,0.57,0.62}
\definecolor{pink4}{rgb}{0.55,0.39,0.42}
\definecolor{pink}{rgb}{1.00,0.75,0.80}
\definecolor{plum1}{rgb}{1.00,0.73,1.00}
\definecolor{plum2}{rgb}{0.93,0.68,0.93}
\definecolor{plum3}{rgb}{0.80,0.59,0.80}
\definecolor{plum4}{rgb}{0.55,0.40,0.55}
\definecolor{plum}{rgb}{0.87,0.63,0.87}
\definecolor{powderblue}{rgb}{0.69,0.88,0.90}
\definecolor{purple1}{rgb}{0.61,0.19,1.00}
\definecolor{purple2}{rgb}{0.57,0.17,0.93}
\definecolor{purple3}{rgb}{0.49,0.15,0.80}
\definecolor{purple4}{rgb}{0.33,0.10,0.55}
\definecolor{purple}{rgb}{0.63,0.13,0.94}
\definecolor{red1}{rgb}{1.00,0.00,0.00}
\definecolor{red2}{rgb}{0.93,0.00,0.00}
\definecolor{red3}{rgb}{0.80,0.00,0.00}
\definecolor{red4}{rgb}{0.55,0.00,0.00}
\definecolor{red}{rgb}{1.00,0.00,0.00}
\definecolor{rosybrown}{rgb}{0.74,0.56,0.56}
\definecolor{royalblue}{rgb}{0.25,0.41,0.88}
\definecolor{saddlebrown}{rgb}{0.55,0.27,0.07}
\definecolor{salmon1}{rgb}{1.00,0.55,0.41}
\definecolor{salmon2}{rgb}{0.93,0.51,0.38}
\definecolor{salmon3}{rgb}{0.80,0.44,0.33}
\definecolor{salmon4}{rgb}{0.55,0.30,0.22}
\definecolor{salmon}{rgb}{0.98,0.50,0.45}
\definecolor{sandybrown}{rgb}{0.96,0.64,0.38}
\definecolor{seagreen}{rgb}{0.18,0.55,0.34}
\definecolor{seashell1}{rgb}{1.00,0.96,0.93}
\definecolor{seashell2}{rgb}{0.93,0.90,0.87}
\definecolor{seashell3}{rgb}{0.80,0.77,0.75}
\definecolor{seashell4}{rgb}{0.55,0.53,0.51}
\definecolor{seashell}{rgb}{1.00,0.96,0.93}
\definecolor{sienna1}{rgb}{1.00,0.51,0.28}
\definecolor{sienna2}{rgb}{0.93,0.47,0.26}
\definecolor{sienna3}{rgb}{0.80,0.41,0.22}
\definecolor{sienna4}{rgb}{0.55,0.28,0.15}
\definecolor{sienna}{rgb}{0.63,0.32,0.18}
\definecolor{skyblue}{rgb}{0.53,0.81,0.92}
\definecolor{slateblue}{rgb}{0.42,0.35,0.80}
\definecolor{slategray}{rgb}{0.44,0.50,0.56}
\definecolor{slategrey}{rgb}{0.44,0.50,0.56}
\definecolor{snow1}{rgb}{1.00,0.98,0.98}
\definecolor{snow2}{rgb}{0.93,0.91,0.91}
\definecolor{snow3}{rgb}{0.80,0.79,0.79}
\definecolor{snow4}{rgb}{0.55,0.54,0.54}
\definecolor{snow}{rgb}{1.00,0.98,0.98}
\definecolor{springgreen}{rgb}{0.00,1.00,0.50}
\definecolor{steelblue}{rgb}{0.27,0.51,0.71}
\definecolor{tan1}{rgb}{1.00,0.65,0.31}
\definecolor{tan2}{rgb}{0.93,0.60,0.29}
\definecolor{tan3}{rgb}{0.80,0.52,0.25}
\definecolor{tan4}{rgb}{0.55,0.35,0.17}
\definecolor{tan}{rgb}{0.82,0.71,0.55}
\definecolor{thistle1}{rgb}{1.00,0.88,1.00}
\definecolor{thistle2}{rgb}{0.93,0.82,0.93}
\definecolor{thistle3}{rgb}{0.80,0.71,0.80}
\definecolor{thistle4}{rgb}{0.55,0.48,0.55}
\definecolor{thistle}{rgb}{0.85,0.75,0.85}
\definecolor{tomato1}{rgb}{1.00,0.39,0.28}
\definecolor{tomato2}{rgb}{0.93,0.36,0.26}
\definecolor{tomato3}{rgb}{0.80,0.31,0.22}
\definecolor{tomato4}{rgb}{0.55,0.21,0.15}
\definecolor{tomato}{rgb}{1.00,0.39,0.28}
\definecolor{turquoise1}{rgb}{0.00,0.96,1.00}
\definecolor{turquoise2}{rgb}{0.00,0.90,0.93}
\definecolor{turquoise3}{rgb}{0.00,0.77,0.80}
\definecolor{turquoise4}{rgb}{0.00,0.53,0.55}
\definecolor{turquoise}{rgb}{0.25,0.88,0.82}
\definecolor{violetred}{rgb}{0.82,0.13,0.56}
\definecolor{violet}{rgb}{0.93,0.51,0.93}
\definecolor{wheat1}{rgb}{1.00,0.91,0.73}
\definecolor{wheat2}{rgb}{0.93,0.85,0.68}
\definecolor{wheat3}{rgb}{0.80,0.73,0.59}
\definecolor{wheat4}{rgb}{0.55,0.49,0.40}
\definecolor{wheat}{rgb}{0.96,0.87,0.70}
\definecolor{whitesmoke}{rgb}{0.96,0.96,0.96}
\definecolor{white}{rgb}{1.00,1.00,1.00}
\definecolor{yellow1}{rgb}{1.00,1.00,0.00}
\definecolor{yellow2}{rgb}{0.93,0.93,0.00}
\definecolor{yellow3}{rgb}{0.80,0.80,0.00}
\definecolor{yellow4}{rgb}{0.55,0.55,0.00}
\definecolor{yellowgreen}{rgb}{0.60,0.80,0.20}
\definecolor{yellow}{rgb}{1.00,1.00,0.00}